\documentclass[aps,prb,twocolumn,
				groupedaddress,superscriptaddress,
				amsfonts,amssymb,amsmath,
				a4paper]{revtex4-1}

\usepackage{amsmath,amssymb,graphicx,xcolor,bm}

\usepackage{hyperref}
\hypersetup{
	colorlinks,
	linkcolor={blue!75!black!80!yellow},
	citecolor={blue!75!black!80!yellow},
	urlcolor={blue!75!black!80!yellow}
}

\begin{document}
\title{Nonlocal optical response in metallic nanostructures}
\date{\today}

\author{S\o ren Raza}
\affiliation{Department of Photonics Engineering, Technical University of Denmark, DK-2800 Kgs. Lyngby, Denmark}
\affiliation{Center for Nanostructured Graphene (CNG), Technical University of Denmark, DK-2800 Kgs. Lyngby, Denmark}
\affiliation{Department of Micro- and Nanotechnology, Technical University of Denmark, DK-2800 Kgs. Lyngby, Denmark}
\author{Sergey I. Bozhevolnyi}
\affiliation{Department of Technology and Innovation, University of Southern Denmark, DK-5230 Odense, Denmark}
\author{Martijn Wubs}
\affiliation{Department of Photonics Engineering, Technical University of Denmark, DK-2800 Kgs. Lyngby, Denmark}
\affiliation{Center for Nanostructured Graphene (CNG), Technical University of Denmark, DK-2800 Kgs. Lyngby, Denmark}
\author{N. Asger Mortensen}
\email[E-mail: ]{asger@mailaps.org}
\affiliation{Department of Photonics Engineering, Technical University of Denmark, DK-2800 Kgs. Lyngby, Denmark}
\affiliation{Center for Nanostructured Graphene (CNG), Technical University of Denmark, DK-2800 Kgs. Lyngby, Denmark}

\begin{abstract}
This review provides a broad overview of the studies and effects of nonlocal response in metallic nanostructures. In particular, we thoroughly present the nonlocal hydrodynamic model and the recently introduced generalized nonlocal optical response (GNOR) model. The influence of nonlocal response on plasmonic excitations is studied in key metallic geometries, such as spheres and dimers, and we derive new consequences due to the GNOR model. Finally, we propose several trajectories for future work on nonlocal response, including experimental setups that may unveil further effects of nonlocal response.
\end{abstract}


\maketitle

\section{Introduction}
The excitation of surface plasmons (SPs), i.e., the collective movement of conduction-band electrons tightly bound to a metal-insulator interface, governs most of the phenomena observed in plasmonic studies. A few examples of the vast number of surface-plasmon related effects feature the large enhancement of the electric field in metal nanoparticles of close proximity~\cite{Hao:2004} and sharp metal geometries,~\cite{Stockman:2004} the squeezing of light beyond the diffraction limit,~\cite{Gramotnev:2010,Gramotnev:2014} and the tunability of the optical properties of metallic structures with size and shape.~\cite{LizMarzan:2006} These appealing properties have found applications in different fields, such as surface-enhanced Raman spectroscopy,~\cite{Moskovits:1985} biosensing,~\cite{Homola:1999,Anker:2008} plasmonic waveguiding,~\cite{Bozhevolnyi:2006b} cancer therapy,~\cite{Lal:2008} and on-chip circuitry.~\cite{Zia:2006a,Huang:2014}

The theoretical modelling of plasmonic phenomena is for the most part based on the macroscopic Maxwell's equations. In particular, the optical response of metals is described through the constitutive relations, which relate the response of the material to the applied field. In the linear regime, the constitutive relation relating the displacement field $\mathbf D$ to the electric field $\mathbf E$ is
\begin{equation}
	\mathbf D(\mathbf r,\omega) = \varepsilon_0 \int \text d\mathbf r' \varepsilon(\mathbf r, \mathbf r',\omega) \mathbf E(\mathbf r',\omega), \label{eq:constitutive_nl}
\end{equation}
where $\varepsilon_0$ is the vacuum permittivity and $\varepsilon(\mathbf r,\mathbf r',\omega)$ denotes the nonlocal permittivity of the metal, here assumed scalar. In a \emph{homogeneous} medium, the nonlocal permittivity depends spatially on $\mathbf r - \mathbf r'$, turning Eq.~\eqref{eq:constitutive_nl} into a convolution which in $\mathbf k$-space becomes the product
\begin{equation}
	\mathbf D(\mathbf k,\omega) = \varepsilon_0 \varepsilon(\mathbf k,\omega) \mathbf E(\mathbf k,\omega). \label{eq:constitutive_nl_k}
\end{equation}
We see that nonlocal response corresponds to a $\mathbf k$-dependent dielectric function. For an isotropic response, we find dependence only on the length of $\mathbf k$, and not its direction.

\begin{figure}
	\centering
	\includegraphics[scale=1]{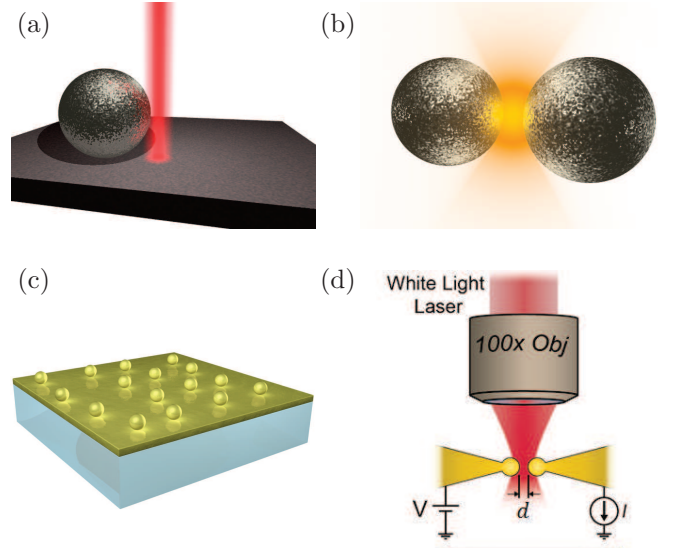}
	\caption{Schematic illustrations of (a) electron beam interacting with silver nanoparticle on thin substrate,~\cite{Ouyang:1992,Scholl:2012,Raza:2013a,Raza:2013c} (b) electromagnetic interaction of two nearby metal nanoparticles,~\cite{Kern:2012,Scholl:2013,Cha:2014,Tan:2014,Kravtsov:2014,Zhu:2014,Hajisalem:2014} (c) gold nanoparticles in close proximity of gold film,~\cite{Ciraci:2012} and (d) optical measurements of two gold-coated atomic-force microscopy tips in close proximity.~\cite{Savage:2012} (c) and (d) courtesy of C. Cirac\`{\i} and J. Baumberg, respectively.}
	\label{fig:fig1}
\end{figure}
In the simpler local-response approximation (LRA), nonlocal effects are neglected and we write $\varepsilon(\mathbf r,\mathbf r',\omega) = \delta(\mathbf r-\mathbf r')\varepsilon(\omega)$, which allows us to straightforwardly perform the integral in Eq.~\eqref{eq:constitutive_nl} as
\begin{equation}
	\mathbf D(\mathbf r,\omega) = \varepsilon_0 \varepsilon(\omega) \mathbf E(\mathbf r,\omega). \label{eq:constitutive_LRA}
\end{equation}
Here, $\varepsilon(\omega)$ is the spatially constant permittivity of the metal, usually described by the Drude-like dielectric function~\cite{Johnson:1972,Rakic:1998a,Hao:2007}
\begin{equation}
	\varepsilon(\omega) = \varepsilon_\infty(\omega) - \frac{\omega_\text{p}^2}{\omega(\omega+i\gamma)}. \label{eq:Drude}
\end{equation}
In Eq.~\eqref{eq:Drude}, $\omega_\text{p}^2 = n_0 e^2/(\varepsilon_0 m)$ is the unscreened plasma frequency of the metal with $n_0$ denoting the equilibrium electron density of the free electrons, $\gamma$ is the Drude damping rate, and $\varepsilon_\infty(\omega)$ is the response from the bound ions and electrons, which accounts for effects such as interband transitions. The importance of the LRA is accentuated by its prevalence in the plasmonic community, being the most commonly applied constitutive description.~\cite{Maier:2007} The LRA has successfully described a plethora of plasmonic phenomena and experiments, such as optical far-field measurements,~\cite{Aizpurua:2003,Abajo:2007,Ergin:2010} electron energy-loss spectroscopy (EELS),~\cite{Ritchie:1957,Nelayah:2007,Bosman:2007,Schaffer:2009,Koh:2011,Nicoletti:2011,Husnik:2013,Raza:2014a} cathodoluminescence experiments,~\cite{Hofmann:2007,Kuttge:2009,Abajo:2010} near-field microscopy,~\cite{Schnell:2009} and surprisingly, even effects in the two-dimensional material graphene~\cite{Chen:2012} and plasmonic particles interspaced by nanometer-sized gaps.~\cite{Duan:2012}

Despite its success, the LRA has been challenged on a number of accounts. One example is the size-dependent SP linewidth broadening observed in metal clusters and small nanoparticles,~\cite{Kreibig:1985,Kreibig:1995,Baida:2009} which has to be phenomenologically accounted for in the LRA through an increased damping rate as a consequence of surface screening and scattering.~\cite{Kreibig:1969,Apell:1983,Uskov:2014} Size-dependent resonance shifts of the SP in noble metal nanoparticles have also been observed in both optical measurements~\cite{Tiggesbaumker:1993,Charle:1998} and EELS [Fig.~\ref{fig:fig1}(a)].~\cite{Ouyang:1992,Scholl:2012,Raza:2013a,Raza:2013c} Another example is the multipole plasmon which, besides the usual surface-plasmon polariton, can be supported by the simple geometry of a metal-vacuum interface~\cite{Tsuei:1990} as a direct consequence of the spill-out of free electrons beyond the classical metal boundary.~\cite{Bennett:1970,Boardman:1975,Bochterle:2012} Thin metal films have also been shown to support resonant excitations above the plasma frequency due to confined longitudinal waves,~\cite{Lindau:1970,Anderegg:1971} which are not taken into account in the LRA.
Recently, several experiments on metal dimers with particles in mutual subnanometer proximity have shown plasmonic effects clearly going beyond the LRA [Figs.~\ref{fig:fig1}(b-d)].~\cite{Ciraci:2012,Savage:2012,Kern:2012,Scholl:2013,Cha:2014,Tan:2014,Kravtsov:2014,Zhu:2014,Hajisalem:2014} A theoretical description of the metal based on ab initio approaches such as density-functional theory (DFT)~\cite{Onida:2002,ZhangP:2014} or similar theories~\cite{Guidez:2014} seem to capture all of the observed non-classical effects.~\cite{Lerme:2010,Teperik:2013a,Andersen:2012,Zuloaga:2009} However, due to the computational demand of such approaches, only very small system sizes (few nanometers) can be considered,~\cite{Teperik:2013a} which puts serious constraints on the feasibility of these approaches for a generic plasmonic system, which is usually tens of nanometers or more. Another simpler and computationally less demanding path is to go beyond the LRA by taking into account nonlocal response through a hydrodynamic approach.~\cite{Boardman:1982a} The hydrodynamic approach has been able to describe size-dependent resonance shifts of noble metal nanoparticles and gap-dependent resonance shift in a particle-film system,~\cite{Ciraci:2012} and can now with the inclusion of electron diffusion~\cite{Mortensen:2014} also describe size-dependent damping and the optical spectra of closely-spaced dimers. Besides taking into account retardation effects (in contrast to DFT) and being physically transparent, significant analytical progress is also possible with the hydrodynamic approach. Many of these properties in the hydrodynamic approach are beneficial in the theoretical studies of generic plasmonic systems with large ($>10$~nm) feature sizes.

The aim of this Topical Review is to give a comprehensive overview of the real-space formulation of nonlocal response in Maxwell's equations. In particular, we review the hydrodynamic~\cite{Boardman:1982a} and the generalized nonlocal optical response (GNOR) models,~\cite{Mortensen:2014} which are two examples of nonlocal response theories. We show that nonlocal response in metals manifests itself through the presence of longitudinal waves. The influence of nonlocal response in metallic nanostructures is exemplified by considering a silver nanosphere, a silver dimer with {\AA}ngstrom-sized gaps, and a core-shell cylinder with a nanometer-thin silver shell. Finally, we discuss future theoretical and experimental directions to unveil the effects of nonlocal response.

\section{Phenomenological theory of nonlocal response}
We begin by considering the real-space formulation of Maxwell's equations taking into account nonlocal response. In the absence of external sources and using Eq.~\eqref{eq:constitutive_nl}, the nonlocal wave equation with assumed scalar nonlocal response reads
\begin{equation}
	\mathbf \nabla \times \mathbf \nabla \times \mathbf E(\mathbf r,\omega) = \left(\frac{\omega}{c}\right)^2 \int \text d\textbf r' \varepsilon(\mathbf r,\mathbf r',\omega) \mathbf E(\mathbf r',\omega), \label{eq:waveequation_gen}
\end{equation}
where we have furthermore assumed the material to be non-magnetic, i.e., $\mathbf B(\mathbf r,\omega) = \mu_0 \mathbf H(\mathbf r,\omega)$, which is applicable for plasmonic metals. We note that in Eq.~\eqref{eq:waveequation_gen} the electromagnetic response of the material is described only through the displacement field $\mathbf D(\mathbf r,\omega)$, which accounts for effects both from the bound and free charges of the metal. Recognizing that the LRA accounts successfully for many of the observed plasmonic phenomena, we proceed by writing the nonlocal permittivity as~\cite{Mortensen:2013,Ginzburg:2013a}
\begin{equation}
	\varepsilon(\mathbf r,\mathbf r',\omega) = \varepsilon(\omega)\delta(\mathbf r-\mathbf r') + f(\mathbf r-\mathbf r',\omega), \label{eq:epsilon_nl}
\end{equation}
where $f(\mathbf r-\mathbf r',\omega)$ is the scalar nonlocal response function associated with a homogeneous medium. We make the justifiable assumptions that $f(\mathbf r-\mathbf r',\omega)$ is symmetric and short-ranged.~\cite{Mortensen:2013} We can express these assumptions mathematically through the moments of the function
\begin{subequations} \label{eq:fproperty}
	\begin{align}
	&\int \text d\mathbf r f(\mathbf r,\omega) \mathbf r = 0, \label{eq:f_property2} \\
	&\int \text d\mathbf r f(\mathbf r, \omega) \mathbf r^2 = 2\xi^2, \label{eq:f_property3}
	\end{align}
\end{subequations}
where $\mathbf r^2 \equiv (x^2,y^2,z^2)$ and we have introduced the length scale $\xi$ as the range (or width) of the nonlocal response function. Since the response function $f(\mathbf r-\mathbf r',\omega)$ is short-ranged [i.e., $\xi$ in Eq.~\eqref{eq:f_property3} is small on the length scale of the variations of the electric field], we Taylor expand the electric field in the integrand of Eq.~\eqref{eq:waveequation_gen} around $\mathbf r$. To second order we find (suppressing the frequency dependency)
\begin{equation}
\begin{aligned}
	E_i(\mathbf{r'}) &\simeq E_i(\mathbf r) + \left[\nabla E_i(\mathbf r) \right] \cdot (\mathbf r' - \mathbf r) \\
	&+ \frac 12 (\mathbf r'- \mathbf r)^T \cdot \left[\mathbf{\hat{H}}E_i(\mathbf r)\right] \cdot (\mathbf r'-\mathbf r), \label{eq:taylor_Efield}
\end{aligned}
\end{equation}
where the Hessian matrix $\mathbf{\hat{H}}$ has the elements $H_{ij} = \partial^2/(\partial_i\partial_j)$ with $i,j=x,y,z$. With these considerations in mind and using the assumptions in Eqs.~(\ref{eq:fproperty}-\ref{eq:taylor_Efield}), we can perform the integral in Eq.~\eqref{eq:waveequation_gen} and find
\begin{equation}
	\mathbf \nabla \times \mathbf \nabla \times \mathbf E(\mathbf r,\omega) = \left(\frac{\omega}{c}\right)^2 \left[ \varepsilon(\omega) + \xi^2 \nabla^2 \right] \mathbf E(\mathbf r,\omega), \label{eq:waveequation_nl}
\end{equation}
where we have absorbed the zeroth-order moment integral of $f(\mathbf r,\omega)$ [i.e., $\int \text d\mathbf r f(\mathbf r,\omega)$] into our definition of $\varepsilon(\omega)$. Interestingly, Eq.~\eqref{eq:waveequation_nl} shows that scalar nonlocal response manifests itself through the Laplacian term, seemingly irrespective of the microscopic or semiclassical origin, and with a strength given by the scalar $\xi$, which relates to the width of the nonlocal response function through Eq.~\eqref{eq:f_property3}. This result also suggests the possibility of several nonlocal mechanisms playing in concert and adding up to an effective $\xi^2$.~\cite{Landau-Lifshitz-Pitaevskii} Furthermore, we have transformed the integro-differential equation of Eq.~\eqref{eq:waveequation_gen} into a regular differential equation, where the presence of the Laplacian operator does not give rise to increased numerical difficulties than the already present double-curl operator.

\section{Hydrodynamic model}\label{sec:hydro}
We discuss now the hydrodynamic model for the free-electron gas, which will allow us to determine the nonlocal length scale $\xi$ introduced in Eq.~\eqref{eq:waveequation_nl}. The idea of modeling the free-electron gas in a hydrodynamic formulation was first introduced by Bloch in a seminal paper in 1933.~\cite{Bloch:1933a} In 1974, Ying~\cite{Ying:1974} extended Bloch's non-retarded approach to a more general density-functional formalism, allowing to go beyond the Thomas--Fermi ground state, which lacked information about the correlation and exchange energies of the electron gas. Shortly after Eguiluz and Quinn~\cite{Eguiluz:1976} included retardation effects. Despite the generalization by Ying,~\cite{Ying:1974} the Thomas--Fermi description of the electron gas in the hydrodynamic model remained popular, and was extensively used in the field of solid-state physics in the 1970s and the beginning of the 1980s. The effect of electron density inhomogeneity and spatial dispersion in planar interfaces,~\cite{Boardman:1975,Boardman:1976a,Melnyk:1970,Eguiluz:1975} multilayered structures,~\cite{Mochan:1987,Mochan:1988} spherical particles~\cite{Boardman:1977,Ruppin:1973,Ruppin:1975} and voids,~\cite{Aers:1979} and cylindrical particles~\cite{Aers:1980,Ruppin:1989} was given a considerable attention. Especially, the results obtained for homogeneous planar surfaces showed good agreement with experiment.~\cite{Boardman:1982a}

Recently, interest in the hydrodynamic model was rekindled when a finite-element numerical implementation of the hydrodynamic equations was presented,~\cite{McMahon:2009,Toscano:2012a,Hiremath:2012} which was subsequently utilized to study the plasmonic cylindrical dimer,~\cite{Toscano:2012a} surface-enhanced Raman spectroscopy,~\cite{Toscano:2012b} and waveguiding in metallic nanostructures.~\cite{Toscano:2013} Simultaneously, application of transformation optics to the hydrodynamic model allowed for analytical solutions of several non-trivial plasmonic structures, even some containing singular geometric features.~\cite{Fernandez-Dominguez:2012a,Fernandez-Dominguez:2012b,Fernandez-Dominguez:2012c} Additionally, numerous different metallic geometries and plasmonic effects have been studied using the hydrodynamic model, such as scattering and mode analysis of cylindrical structures, including nanotubes,~\cite{Villo-Perez:2009,Raza:2011,Yan:2013,Toscano:2013,Raza:2013d,Moradi:2014,Ben:2012} roughness effects on plasmonic tips,~\cite{Wiener:2012,Ruppin:2005} nonlinear effects in nonlocal media,~\cite{Ciraci:2012b,Ciraci:2012c,Li:2014} scattering of light off three-dimensional nanostructures,~\cite{David:2011,Ciraci:2012,Ciraci:2013,Trivedi:2014} surface plasmon propagation in metal-insulator, metal-insulator-metal, insulator-metal-insulator, and hourglass waveguides,~\cite{Moreau:2013,Raza:2013b,Wiener:2013b,Xiao:2014} epsilon-near-zero and perfect imaging effects,~\cite{Larkin:2005,Ceglia:2013,David:2013,Yanai:2014} influence of nonlocal response on the Casimir force,~\cite{Sun:2014} studies of hyperbolic metamaterials and periodic media,~\cite{Yan:2012,Yan:2013b,Yanai:2013,Benedicto:2014} investigations of nonlocal effects in EELS,~\cite{Wiener:2013,Christensen:2014} and coupling of dipole emitters to plasmonic structures.~\cite{Fuchs:1981,Christensen:2014,Filter:2014} Theoretical work has also been done to compare the hydrodynamic approach with more advanced approaches such as density-functional theory.~\cite{Teperik:2013a,Teperik:2013b,Stella:2013} Finally, other nonlocal models besides the hydrodynamic model have also been utilized to study different geometries.~\cite{Fuchs:1969,Kliewer:1974,Xue:2014,Khurgin:2014,Apell:1982a,Apell:1983,Abajo:2008}

As a detailed derivation of the hydrodynamic model has been reported before,~\cite{Boardman:1982a,Bloch:1933a,Eguiluz:1976,Pitarke:2007} we focus here on the most important steps of the derivation and extract the essential physics of the hydrodynamic model. The basic assumption of the hydrodynamic model is that the many-electron energy and dynamics can be expressed in terms of a scalar field and a velocity field, the electron density $n(\mathbf r,t)$ and the hydrodynamic velocity $\mathbf v(\mathbf r,t)$, respectively. The energy of the electron plasma is a functional of the electron density and velocity. The dynamics of these variables under the influence of macroscopic electromagnetic fields $\mathbf E(\mathbf r,t)$ and $\mathbf B(\mathbf r,t)$ is obtained by functional differentiation of the energy (i.e., Hamilton's principle). Functional differentiation with respect to the velocity field gives the hydrodynamic equation of motion~\cite{Boardman:1982a}
\begin{equation}
	\left[ \partial_t + \mathbf v \cdot \mathbf \nabla \right] \mathbf v =  -\frac{e}{m} \left[\mathbf E + \mathbf v \times \mathbf B \right] - \mathbf \nabla \frac{\delta G[n]}{\delta n} - \gamma \mathbf v, \label{eq:hydro_functional}
\end{equation}
while functional differentiation with respect to the electron density gives the continuity equation
\begin{equation}
	\partial_t n = - \mathbf \nabla \cdot \left( n \mathbf v \right), \label{eq:hydro_continuity}
\end{equation}
expressing charge conservation. On the right-hand side of Eq.~\eqref{eq:hydro_functional}, the first term is the Lorentz force, while the second term can take into account the correlation, exchange and the internal kinetic energy of the electron gas, if an appropriate functional $G[n]$ is chosen.~\cite{Ying:1974} Here, $\delta G[n]/\delta n$ denotes the functional derivative. The last term of Eq.~\eqref{eq:hydro_functional}, which represents damping in terms of the bulk damping rate $\gamma$, cannot be obtained from the energy functional approach and is therefore added phenomenologically.

The most common, and also the simplest, approach~\cite{Boardman:1982a,Bennett:1970} is to use the Thomas--Fermi model for the functional $G[n]$, given as
\begin{equation}
	G[n(\mathbf r,t)] = \int \text d\mathbf r \frac{3h^2}{10m} \left(\frac{3}{8\pi}\right)^{\frac 23} n^{\frac 53}(\mathbf r,t), \label{eq:TF_functional}
\end{equation}
which describes only the internal kinetic energy of the electron gas. The functional derivative of Eq.~\eqref{eq:TF_functional} can now be performed
\begin{equation}
	\frac{\delta G[n]}{\delta n} = \frac{h^2}{2m}  \left(\frac{3}{8\pi}\right)^{\frac 23} n^{\frac 23}(\mathbf r,t), \label{eq:TF_functional_der}
\end{equation}
which upon insertion in Eq.~\eqref{eq:hydro_functional} finally gives
\begin{equation}
	\left[ \partial_t + \mathbf v \cdot \mathbf \nabla \right] \mathbf v = -\frac{e}{m} \left[\mathbf E + \mathbf v \times \mathbf B \right] - \frac{\beta^2}{n} \mathbf \nabla n - \gamma \mathbf v. \label{eq:hydro_pressure}
\end{equation}
In analogy with fluid hydrodynamics, the gradient term on the right-hand side of Eq.~\eqref{eq:hydro_pressure} represents the pressure, which is proportional to $\beta^2=3/5 v_\textsc{f}^2$ with $v_\textsc{f}$ denoting the Fermi velocity,~\cite{Halevi:1995} and describes a force that will act to homogenize any inhomogeneity in the electron density. The pressure term gives rise to nonlocal response in the hydrodynamic model.

In the spirit of linear-response theory, we now follow the usual approach~\cite{Boardman:1982a,Pitarke:2007} to solve Eqs.~\eqref{eq:hydro_continuity} and \eqref{eq:hydro_pressure}, by expanding the physical fields in a static term (e.g., $n_0$ is the homogeneous static electron density), and a small (by assumption) first-order dynamic term, as in perturbation theory, thereby linearizing the equation of motion and the continuity equation. In the frequency domain, we obtain the coupled electromagnetic equations~\cite{Raza:2011,Toscano:2012a}
\begin{subequations}
	\label{eq:hydro_coupled}
	\begin{align}
	&\mathbf \nabla \times \mathbf \nabla \times \mathbf E(\mathbf r,\omega) = \left( \frac{\omega}{c} \right)^2 \varepsilon_\infty \mathbf E(\mathbf r,\omega) + i\omega\mu_0 \mathbf J(\mathbf r,\omega), \label{eq:hydro_waveequation} \\	
	&\frac{\beta^2}{\omega\left(\omega+i\gamma\right)} \mathbf \nabla \left[ \mathbf \nabla \cdot \mathbf J(\mathbf r,\omega) \right] + \mathbf J(\mathbf r,\omega) = \sigma \mathbf E(\mathbf r,\omega), \label{eq:hydro_linear}
	\end{align}
\end{subequations}
where $\mathbf J(\mathbf r,\omega) = -en_0\mathbf v(\mathbf r,\omega)$ is the induced current density, and $\sigma=\varepsilon_0 i \omega_\text{p}^2/(\omega + i\gamma)$ is the Drude conductivity which relates to the Drude permittivity [Eq.~\eqref{eq:Drude}] as $\varepsilon(\omega) = \varepsilon_\infty + i\sigma/(\varepsilon_0 \omega)$. We see that in the LRA limit of $\beta \rightarrow 0$, Eq.~\eqref{eq:hydro_linear} reduces to Ohm's law. Now, by combining Eqs.~\eqref{eq:hydro_waveequation} and \eqref{eq:hydro_linear}, we can rewrite the governing equations in the hydrodynamic model as
\begin{equation}
	\mathbf \nabla \times \mathbf \nabla \times \mathbf E(\mathbf r,\omega) = \left(\frac{\omega}{c}\right)^2 \left[ \varepsilon(\omega) + \xi_\textsc{h}^2 \mathbf \nabla (\mathbf \nabla \cdot) \right] \mathbf E(\mathbf r,\omega), \label{eq:waveequation_hydro}
\end{equation}
where we find the nonlocal parameter in the hydrodynamic model to be~\cite{Toscano:2013,Mortensen:2014}
\begin{equation}
	\xi_\textsc{h}^2(\omega) = \frac{\varepsilon_\infty(\omega) \beta^2}{\omega(\omega+i\gamma)}. \label{eq:xi_hydro}
\end{equation}
In the absence of interband effects and bulk damping mechanisms ($\varepsilon_\infty=1$ and $\gamma=0$), the nonlocal parameter is simply $\xi_\textsc{h} = \beta/\omega$, which is a purely real-valued quantity, i.e., the distance travelled by a conduction electron during an optical cycle. We also see that the nonlocal parameter increases with the Fermi velocity (through $\beta$). In Table~\ref{tab:tab1} we list values for $\omega_\text{p}$, $v_\textsc{f}$, and $\gamma$ for relevant plasmonic metals. In addition, the response from the bound electrons $\varepsilon_\infty(\omega)$ can be determined from the measured bulk dielectric functions $\varepsilon_\text{exp}(\omega)$ using the recipe $\varepsilon_\infty(\omega) = \varepsilon_\text{exp}(\omega) +\omega_\text{p}^2/(\omega^2+i\gamma\omega)$.~\cite{Abajo:2008}

Comparing the hydrodynamic wave equation [Eq.~\eqref{eq:waveequation_hydro}] with the wave equation from the phenomenological nonlocal model [Eq.~\eqref{eq:waveequation_nl}], we see that the mathematical operator responsible for the nonlocal effects is the gradient-of-the-divergence and not the Laplacian. In fact, a closer $\mathbf k$-space analysis of Eq.~\eqref{eq:waveequation_nl} reveals that the Laplacian operator introduces a weak spatial dispersion in the transverse part of the electric field,~\cite{Toscano:2013} which is in contrast with the hydrodynamic wave equation where spatial dispersion only affects the longitudinal part of the electric field.~\cite{Raza:2011} The Laplacian operator in the phenomenological model stems from the \emph{scalar} nonlocal response function in Eq.~\eqref{eq:epsilon_nl}.

Besides the hydrodynamic wave equation in Eq.~\eqref{eq:waveequation_hydro}, an additional enlightening way of writing up the nonlocal electromagnetic equations for the electric field is in terms of the divergence and curl of the electric field~\cite{Boardman:1976a,Raza:2011,Raza:2013b}
\begin{subequations}
	\label{eq:diff_E}
	\begin{align}
	&\left( \nabla^2 + k_\textsc{l}^2 \right) \mathbf \nabla \cdot \mathbf E(\mathbf r,\omega) = 0, \label{eq:diff_divE} \\
	&\left( \nabla^2 + k_\textsc{t}^2 \right) \mathbf \nabla \times \mathbf E(\mathbf r,\omega) = 0.
	\label{eq:diff_curlE}
	\end{align}
\end{subequations}
Here, $k_\textsc{l}^2=\varepsilon(\omega)/\xi_\textsc{h}^2$ and $k_\textsc{t}^2 = (\omega/c)^2 \varepsilon(\omega)$ are the wave vectors of the longitudinal and transversal electric field, respectively. In the Fourier domain ($\mathbf k$-space), it becomes clear that Eq.~\eqref{eq:diff_divE} describes the longitudinal part of the electric field, while Eq.~\eqref{eq:diff_curlE} corresponds to the transversal part of the electric field. 
The longitudinal and tranversal electric fields are two different types of waves, which in a homogeneous medium are uncoupled, but can in the presence of an interface be coupled by means of the electromagnetic boundary conditions. We stress that the difference between the LRA and the inclusion of nonlocal response in the hydrodynamic approach is the presence of the longitudinal wave, which will be responsible for all nonlocal effects.

\subsection{Additional boundary condition}
Within the LRA, Maxwell's boundary conditions, commonly derived using pill-box and current loop arguments, are sufficient to determine the amplitudes of the transversal electric and magnetic fields. However, the presence of an additional wave due to nonlocal response will require an additional boundary condition (ABC) to determine the amplitude of the longitudinal wave.~\cite{Melnyk:1970} Inspired by the many discussions on the appropriate boundary conditions in the nonlocal hydrodynamic model,~\cite{Boardman:1982a,Melnyk:1970,Sauter:1967,Forstmann:1977,Boardman:1981a,Jewsbury:1981} we will assume that the static equilibrium free-electron density $n_0$ of the metal has a step profile, i.e., $n_0$ is constant inside the metal and abruptly drops to zero at the metal-dielectric interface. The consequence of this assumption is that the induced charge density $\rho$ will vanish at the metal-dielectric boundary, at which a pill-box argument on the continuity equation $\mathbf \nabla \cdot \mathbf J = i\omega \rho$ reveals the additional boundary condition
\begin{equation}
\mathbf J \cdot \mathbf{\hat{n}} = 0, \label{eq:ABC}
\end{equation}
stating that the normal component of the induced current density $\mathbf J$ vanishes at the metal boundary. The step profile of $n_0$ is also at the heart of the LRA, but in contrast to Eq.~\eqref{eq:ABC}, the normal component of the induced current at the metal boundary will in general not vanish, due to the LRA constitutive relation $\mathbf J(\mathbf r,\omega) = \sigma(\omega) \mathbf E(\mathbf r,\omega)$. We note that the ABC in Eq.~\eqref{eq:ABC} will not allow us to include the quantum mechanical effect of spill-out of electrons occurring due to the finite potential difference at the metal-dielectric interface. In fact, the ABC corresponds to the assumption of an infinite work function in more microscopic theories.

We add that the ABC stated in Eq.~\eqref{eq:ABC}, which is valid for a metal-dielectric interface, can be rewritten in terms of the normal components of the electric field as~\cite{Boardman:1982a,Toscano:2013}
\begin{equation}
\varepsilon_\infty \mathbf E_\text{m} \cdot \mathbf{\hat{n}} = \varepsilon_\text{d} \mathbf E_\text{d} \cdot \mathbf{\hat{n}}, \label{eq:ABC2}
\end{equation}
where $\varepsilon_\text{d}$ is the permittivity of the dielectric, $\varepsilon_\infty$ is the response due to the bound charges in the metal [cf. Eq.~\eqref{eq:Drude}], and $\mathbf E_\text{m}$ and $\mathbf E_\text{d}$ are the electric fields in the metal and dielectric, respectively. Thus, in the special case of a metal-vacuum interface with a metal that has no response due to bound charges ($\varepsilon_\infty=1$), then the normal component of the electric field is continuous. For completeness, we add that a second ABC is needed at a metal-metal interface,\cite{Forstmann:1978,Boardman:1981a,Jewsbury:1981} but we will not discuss such interfaces in the following.

\section{Generalized nonlocal optical response}\label{sec:gnor}
A hitherto disregarded effect in the discussion of metallic nanostructures and nonlocal response is the classical phenomenon of electron diffusion.~\cite{Hanson:2010,Landau-Lifshitz-Pitaevskii} While the hydrodynamic model incorporates the convective current due to the pressure term in Eq.~\eqref{eq:hydro_pressure}, it completely neglects any currents due to diffusion. The GNOR model expands the hydrodynamic theory to also take into account electron diffusion. We now consider the mathematical description of this effect. The inclusion of electron diffusion alters the continuity equation, which in its linearized form now reads
\begin{equation}
	\begin{aligned}
		-i\omega e n(\mathbf r,\omega) &= D \nabla^2 [e n(\mathbf r,\omega)] + \mathbf \nabla \cdot [-en_0 \mathbf v(\mathbf r,\omega)] \\
		&= \mathbf \nabla \cdot \mathbf J(\mathbf r,\omega), \label{eq:convection_diffusion}
	\end{aligned}
\end{equation}
also known as the convection-diffusion equation. Here, $D$ is the diffusion constant, and the induced current density, given by Fick's law, now has a diffusive contribution
\begin{equation}
	\mathbf J(\mathbf r,\omega) = -en_0 \mathbf v(\mathbf r,\omega) + e D \mathbf \nabla n(\mathbf r,\omega). \label{eq:current_Fick}
\end{equation}
Combining the convection-diffusion equation and Fick's law for the current density with the linearized form of the hydrodynamic equation [Eq.~\eqref{eq:hydro_pressure}],~\cite{Mortensen:2014} we eventually arrive at the following constitutive relation for the current density
\begin{equation}
	\left[\frac{\beta^2}{\omega(\omega+i\gamma)} + \frac{D}{i\omega} \right] \mathbf \nabla [\mathbf \nabla \cdot \mathbf J(\mathbf r,\omega)] + \mathbf J(\mathbf r,\omega) = \sigma \mathbf J(\mathbf r,\omega), \label{eq:gnor_linear}
\end{equation}
which we immediately recognize to have the same form as the hydrodynamic constitutive relation [Eq.~\eqref{eq:hydro_linear}]. The difference lies in the prefactor of the first term, which we can rewrite as
\begin{equation}
	\frac{\beta^2}{\omega(\omega+i\gamma)} + \frac{D}{i\omega} = \frac{\beta^2+D(\gamma-i\omega)}{\omega(\omega+i\gamma)} \equiv \frac{\eta^2}{\omega(\omega+i\gamma)},
\end{equation}
where we have defined the parameter
\begin{equation}
	\eta^2 \equiv \beta^2 + D(\gamma - i\omega). \label{eq:eta_nl}
\end{equation}
Comparing Eq.~\eqref{eq:gnor_linear} with Eq.~\eqref{eq:hydro_linear}, we see that the mathematical considerations from the hydrodynamic model can be mapped directly to the GNOR model using the simple substitution $\beta^2 \rightarrow \eta^2$. Thus, we straightforwardly find the GNOR nonlocal parameter to be
\begin{equation}
	\xi_\textsc{gnor}^2 = \frac{\varepsilon_\infty(\omega) [\beta^2 + D(\gamma - i\omega)]}{\omega(\omega+i\gamma)}. \label{eq:xi_gnor}
\end{equation}
and similarly, the longitudinal wave vector in the GNOR model is $k_\textsc{l}^2 = \varepsilon(\omega)/\xi_\textsc{gnor}^2$.

Considering the specific case with no interband effects and no bulk damping, we see that the diffusion constant only contributes to the imaginary part of $\xi_\textsc{gnor}^2$, making the nonlocal parameter complex-valued (in contrast to the hydrodynamic model). Additionally, $\beta^2$ contributes only to the real part of $\xi_\textsc{gnor}^2$ (as in the hydrodynamic model). In the general case, the hydrodynamic parameter $\beta \propto v_\textsc{f}$ (or the convective current) is still the main contributor to the real part of $\xi_\textsc{gnor}$, while the imaginary part of $\xi_\textsc{gnor}$ is mostly characterized by the diffusion constant $D$ (or the diffusive current). Table~\ref{tab:tab1} lists values for $D$ for relevant plasmonic metals.

A more elaborate description of the response of the free electrons in metals based on the Boltzmann equation has been reported by Lindhard,~\cite{Lindhard:1954} which includes Landau damping but not diffusive currents. Inclusion of diffusion in the Boltzmann description is possible by ensuring the perturbed electron density to relax to the local electron density~\cite{Warren:1960} (rather than to the unperturbed electron density as done in the Lindhard description). This correction to the Boltzmann dielectric function was provided by Mermin.~\cite{Mermin:1970} Interestingly, Halevi~\cite{Halevi:1995} compared the dielectric function of the hydrodynamic model with the Boltzmann--Mermin dielectric function in the small $k$-limit and showed that
\begin{equation}
	\eta_\text{Halevi}^2 = \frac{\tfrac 35 \omega +i \tfrac 13 \gamma}{\omega+i\gamma} v_\textsc{f}^2, \label{eq:eta_Halevi}
\end{equation}
to ensure agreement between the two models, which is in contrast to the usual $\eta = \beta$ in the hydrodynamic model. In particular, the result by Halevi shows that the Mermin correction (i.e., inclusion of diffusion in the Boltzmann equation) renders the usual hydrodynamic $\beta$ parameter complex-valued, exactly as in the GNOR theory where the substitution $\beta^2\rightarrow\eta^2$ is crucial. Comparison of the imaginary parts of Eqs.~\eqref{eq:eta_nl} and \eqref{eq:eta_Halevi} provides an estimate for the diffusion constant $D$ as
\begin{equation}
	D = \frac{4}{15} \frac{\gamma}{\omega^2+\gamma^2} v_\textsc{f}^2. \label{eq:D_Halevi}
\end{equation}
Inserting appropriate values for the different parameters in Eq.~\eqref{eq:D_Halevi}, see Table~\ref{tab:tab1}, we find that $D\approx 10^{-6}$~m$^2$/s at optical frequencies, which is approximately two orders of magnitude smaller than the values for $D$ for nanoparticles shown in Table~\ref{tab:tab1}. This discrepancy of values of the diffusion constant shows that the diffusion in plasmonic nanoparticles cannot be explained by bulk diffusion. On the theoretical side, it is important to see that inclusion of diffusion in the Boltzmann equation leads to a complex-valued $\eta$ just as the inclusion of diffusion in the hydrodynamic model (i.e., GNOR theory) also results in a complex-valued $\eta$. Observed diffusion constants in nanoparticles are thus mainly due to surface effects, but we are not aware of a derivation based on the Boltzmann equation of surface diffusion with the correct magnitude.

\section{Nonlocal effects in plasmonic systems}
This section is devoted to studying some of the relevant plasmonic systems, which exhibit features due to nonlocal response. We begin in Sec.~\ref{sec:sphere} by giving an overview of the electromagnetic response of a single spherical nanoparticle, where the important length scale is the particle diameter. Next in Sec.~\ref{sec:dimer} we consider the plasmonic dimer, consisting in this case of two infinitely long cylinders, and study the dependence of the optical spectrum on the gap size. Finally in Sec.~\ref{sec:coreshell}, we also take a look at the interesting properties of a core-shell nanowire, consisting of an insulating core and a nanometer-sized metallic shell. We will mainly compare the GNOR model with the LRA, while comparison with the hydrodynamic model will also be displayed when relevant.

\subsection{Spherical particle}\label{sec:sphere}
The metallic spherical particle represents an archetypical geometry studied in plasmonics due to the support of localized surface plasmons and the presence of analytical solutions. The electromagnetic scattering problem of a metal sphere of radius $R$ and Drude permittivity $\varepsilon(\omega)$, which is excited by a plane monochromatic wave and homogeneously embedded in a material with dielectric constant $\varepsilon_\textsc{b}$, was first analytically solved by Mie in the LRA.~\cite{Mie:1908} The exact solution provided by Mie is commonly named Mie theory~\cite{vandeHulst:1981,Bohren:1983} and takes into account retardation effects. Later, Ruppin~\cite{Ruppin:1973,Ruppin:1975} extended the Mie theory to include nonlocal response in the metallic sphere by taking into account the longitudinal wave. In the simpler non-retarded limit, the multipolar polarizability of the metal sphere was extended to include nonlocal effects~\cite{Apell:1982a,Apell:1982b,Dasgupta:1981,Fuchs:1987,Crowell:1968} and used to study the optical properties of very small particles $R<10$~nm, where retardation effects can for the most part be safely neglected.
\begin{figure}[!tb]
	\centering
	\includegraphics[width=1\columnwidth]{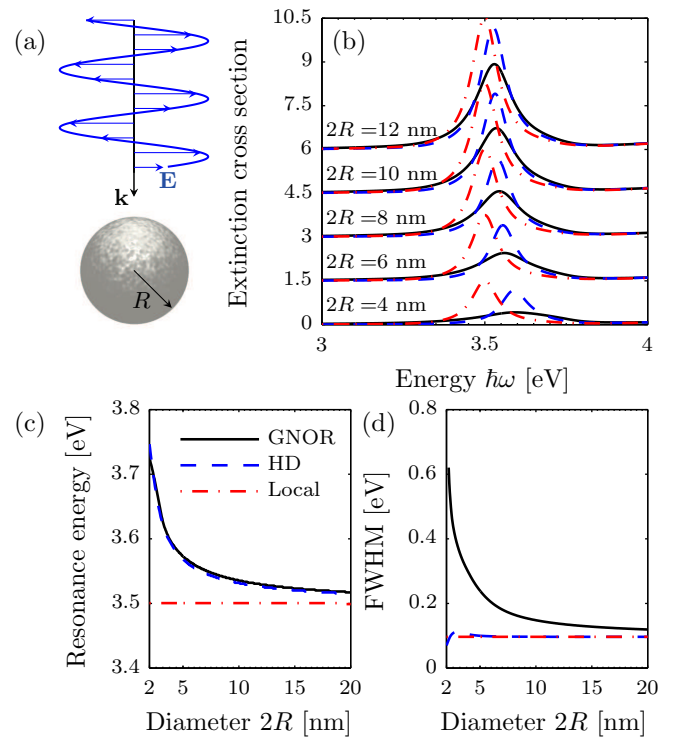}
	\caption{(a) Sketch of a plane wave impinging on a silver sphere of radius $R$. (b) Extinction cross section (in units of the geometrical cross section $\pi R^2$ of the sphere) of a silver sphere in vacuum for decreasing sphere diameter calculated using the GNOR model (black solid lines), hydrodynamic model (blue dashed lines) and LRA (red dash-dotted lines). For clarity, each spectrum is displaced vertically by 1.5 normalized unit. (c-d) Resonance energy and full-width at half maximum (FWHM) of the dipole mode of a silver sphere as a function of diameter $2R$, respectively.}
	\label{fig:fig2a}
\end{figure}

In this section, we present the nonlocal dipolar polarizability, which in the LRA is described by the Clausius--Mossotti factor,~\cite{Maier:2007} to study the extinction cross section of spheres with diameters below $20$~nm. In this size range and under the excitation of a plane wave, the effect of retardation is small and the response of the metal sphere is dominated by the dipolar mode, as we will see in the comparison with fully retarded calculations in Sec.~\ref{sec:retardation}. By considering the poles of the nonlocal dipolar polarizability, we directly show that the nonlocal parameters $\beta$ and $D$ relate to the SP resonance energy and linewidth, respectively.

The derivation of the nonlocal dipolar polarizability $\alpha_\textsc{nl}$ is detailed elsewhere,~\cite{Raza:2013c} so here we simply state the result for the non-retarded limit
\begin{subequations}\label{eq:alphaNL}
	\begin{align}
	\alpha_\textsc{nl}=4\pi R^3 \frac{\varepsilon-\varepsilon_\textsc{b}\left(1+\delta_\textsc{nl}\right)} {\varepsilon+2\varepsilon_\textsc{b}\left(1+\delta_\textsc{nl}\right)},
	\end{align}
where the nonlocal correction is given as
	\begin{align}
	\delta_\textsc{nl}=\frac{\varepsilon-\varepsilon_\infty}{\varepsilon_\infty} \frac{j_1(k_\textsc{l}R)}{k_\textsc{l} R j_1'(k_\textsc{l}R)}.
	\end{align}
\end{subequations}
Here, $j_l$ is the spherical Bessel function of the first kind of angular-momentum order $l$, and the prime denotes differentiation with respect to the argument. In the LRA, $\delta_\textsc{nl}\rightarrow 0$ and Eq.~\eqref{eq:alphaNL} simplifies to the LRA dipolar polarizability, described by the well-known Clausius--Mossotti factor $(\varepsilon-\varepsilon_\textsc{b})/(\varepsilon+2\varepsilon_\textsc{b})$. We note that nonlocal effects enter the Clausius--Mossotti factor as an elegant and simple rescaling of either the metal permittivity~\cite{Apell:1983} from $\varepsilon$ to $\tilde{\varepsilon}= \varepsilon (1+\delta_\textsc{nl})^{-1}$ or of the background permittivity~\cite{Luo:2013} from $\varepsilon_\textsc{b}$ to $\tilde{\varepsilon}_\textsc{b} = \varepsilon_\textsc{b} (1+\delta_\textsc{nl})$.

The SP resonance energy follows theoretically from the Fr\"{o}hlich condition, i.e., we must consider the poles of Eq.~(\ref{eq:alphaNL})
\begin{equation}
	\varepsilon + 2\varepsilon_\textsc{b} \left(1+\delta_\textsc{nl}\right) = 0, \label{eq:alphaNL_pole}
\end{equation}
which will be given by a complex-valued resonance frequency $\omega = \omega'+i\omega''$. The real part $\omega'$ gives the SP resonance frequency, while the imaginary part $\omega''$ is related to the SP resonance linewidth. In the following analytical analysis based on the GNOR model, we consider for simplicity the case of a particle in vacuum ($\varepsilon_\textsc{b}=1$) with no interband effects ($\varepsilon_\infty=1$) and find (to first order in 1/R)~\cite{Mortensen:2014}
\begin{subequations}
	\label{eq:omega_pole}
	\begin{align}
	\omega' &= \frac{\omega_\text{p}}{\sqrt{3}} + \frac{\sqrt{2}\beta}{2R}, \label{eq:omega_real} \\
	\omega'' &= -\frac{\gamma}{2} - \frac{\sqrt{6}}{12} \frac{D\omega_\text{p}}{\beta R}. \label{eq:omega_imag}
	\end{align}
\end{subequations}
In Eq.~\eqref{eq:omega_real}, the first term is the common size-independent local-response Drude result for the SP resonance that also follows from the poles of the LRA polarizability, and the second term gives the size-dependent blueshift due to the hydrodynamic pressure. In Eq.~\eqref{eq:omega_imag}, we have again the size-independent LRA term, while the second term shows a size-dependent linewidth due to diffusion. We can also clearly see that only a size-dependent blueshift is present in the hydrodynamic model, while the GNOR also accounts for a size-dependent SP linewidth. With the inclusion of interband effects, this clear distinction becomes somewhat blurred, since the hydrodynamic model will also show an extremely weak size-dependent linewidth. The origin to the size dependence in nonlocal response is from the smearing of the induced surface charges over a finite distance (few~{\AA}).~\cite{Feibelman:1982,Pendry:2012,Toscano:2012a,Toscano:2012b,Yan:2013} Yan~\cite{Yan:2013} showed explicitly how the smearing of induced charges into the metal leads to a size-dependent blueshift in the hydrodynamic model. In contrast, the delta-function behaviour of the induced surface charges in the LRA leads to no dependence on size. We point out that a $1/R$ dependence on the blueshift and the linewidth of the SP resonance energy of small Ag nanoparticles has been experimentally observed using optical spectroscopy.~\cite{Kreibig:1969,Genzel:1975,Charle:1984,Charle:1989,Charle:1996,Charle:1998,Kreibig:1985,Tiggesbaumker:1993}

With the nonlocal polarizability we can determine the extinction cross section $C_\text{ext}$, which is the sum of the scattering and absorption cross sections, of a metal sphere using the relation~\cite{Bohren:1983}
\begin{align}
	C_\text{ext} = \frac{k_\textsc{b}^4}{6 \pi} |\alpha_\textsc{nl}|^2 + k_\textsc{b} \text{Im}(\alpha_\textsc{nl}), \label{eq:extinction_nr}
\end{align}
where $k_\textsc{b}=(\omega/c)\sqrt{\varepsilon_\textsc{b}}$ is the wave vector in the homogeneous background dielectric medium. Eq.~\eqref{eq:extinction_nr} allows for a more quantitative assessment, based on an observable quantity, of the size-dependent blueshift and linewidth broadening of the dipolar SP resonance anticipated from the approximate analytical relations in Eq.~\eqref{eq:omega_pole}. In Fig.~\ref{fig:fig2a}(b) we show the extinction cross section of a small Ag sphere, sketched in Fig.~\ref{fig:fig2a}(a), with diameter varying from 4~nm to 12~nm, calculated in the GNOR theory (black line), hydrodynamic model (blue line) and LRA (red line). As expected, the LRA shows no change in resonance energy or linewidth of the dipolar SP with size. The hydrodynamic model shows a blueshift of the SP with decreasing diameter and a slightly smaller SP amplitude as a consequence of small, but finite, surface absorption due to bulk material losses and interband transitions. The GNOR model shows both blueshift and linewidth broadening of the SP resonance with decreasing particle size. Quantitatively, we see in Figs.~\ref{fig:fig2a}(c) and (d) that a blueshift of $\sim 0.2$~eV and an increased linewidth broadening of $\sim 0.5$~eV is seen when the sphere diameter decreases from 20~nm to 2~nm. We note that the size-dependent blueshift exceeds the $1/R$-dependency, given by the analytical relations in Eq.~\eqref{eq:omega_pole}, for diameters below $10$~nm and higher-order terms become important.~\cite{Raza:2013a} The hydrodynamic model shows the same blueshift of the SP as the GNOR model, but no significant increase in linewidth. In particular, a weak size-dependent linewidth in the hydrodynamic model can be seen in Fig.~\ref{fig:fig2a}(d) for diameters below $5$~nm, which is due to the inclusion of bulk losses (i.e., $\gamma \neq 0$) and interband effects.

\subsubsection{Size-dependent resonance shift}\label{sec:blueshift}
\begin{figure}[!tb]
	\centering
	\includegraphics[width=1\columnwidth]{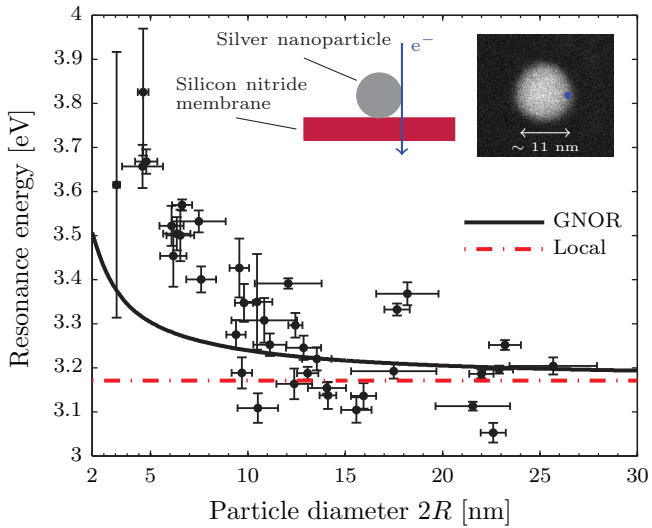}\\
	\caption{Resonance energy as function of particle diameter $2R$ of silver nanoparticles dispersed on a $10$~nm silicon nitride membrane measured with EELS. The EELS data is collected by positioning the electron beam at the surface of the silver particle to strongly excite surface plasmons (see insets).~\cite{Raza:2013a} The curves show calculations of a silver sphere in a homogeneous background using Eq.~\eqref{eq:alphaNL} in the GNOR model (black solid line) and LRA (red dash-dotted line). From the average resonance energy of the largest particles $2R>20$~nm, we fit $\varepsilon_\textsc{b}=1.98$.}
	\label{fig:fig2b}
\end{figure}
The size-dependent shift of the SP resonance energy of small metal particles has been experimentally observed on several occasions.~\cite{Charle:1998,Kreibig:1985,Tiggesbaumker:1993,Ouyang:1992,Zhao:1996,Genzel:1975,Brechignac:1992} It has been found that alkali metals, such as potassium, redshift with decreasing particle size, while noble metals, such as silver and gold, blueshift with decreasing particle size. While the redshift in alkali metals is explained to be a consequence of the spill-out of free electrons, the blueshift in noble metals is mainly attributed to the screening from lower-lying band electrons, such as the $d$-band in silver,~\cite{Liebsch:1993} although additional sources for the blueshift are also present.~\cite{Genzel:1975,Scholl:2013,Monreal:2013}

Recently, EELS measurements on silver nanoparticles dispersed on thin ($\leq 10$~nm) substrates have shown a strong blueshift of the SP resonance energy, when the particle diameter decreases from approximately $25$~nm down to $2$~nm.~\cite{Scholl:2012,Raza:2013a,Raza:2013c} In Fig.~\ref{fig:fig2b} we show EELS measurements of the resonance energy of silver nanoparticles dispersed on a $10$~nm silicon nitride membrane as a function of diameter.~\cite{Raza:2013a} We see that the SP resonance energy shows a significant increase from approximately $3.2$~eV to $3.7$~eV in the diameter range considered. Additionally, we show in Fig.~\ref{fig:fig2b} calculations of the SP resonance energy using the dipolar polarizability [Eq.~\eqref{eq:alphaNL}] in the GNOR model (black line) and in the LRA (red line). As a crude approximation we may account for the effect of the substrate by incorporating it into an effective homogeneous background permittivity $\varepsilon_\textsc{b}$. In this approximation, one uses the average resonance energy of the largest particles $2R>20$~nm to fit the background permittivity $\varepsilon_\textsc{b}$ to ensure agreement with the LRA in the large-diameter range.~\cite{Scholl:2012,Raza:2013a} The assumption behind this approach is that the effective background permittivity does not depend on the radius of the particle, but more detailed theoretical calculations by Yan~\cite{Yan:2013} show variations of 20\% for particle radii varying from 2 to 18 nm. As expected from the discussion of Fig.~\ref{fig:fig2a}, the SP resonance energy in the LRA is size-independent, while the GNOR model shows a blueshift, which is in qualitative agreement with the experiments. However, the experimentally-measured blueshift is larger. This difference was initially conjectured to be related to the inaccurate modeling of the substrate,~\cite{Raza:2013a} however, a proper inclusion of the electromagnetic effects of the thin substrate has interestingly been shown not to be able to account for the discrepancy.~\cite{Raza:2013c} Instead the explanation for the discrepancy may be related to more complicated phenomena in silver, such as an inhomogeneous equilibrium electron density due to Friedel oscillations and electron spill-out~\cite{Raza:2013a} or changes in the electronic band structure.~\cite{Genzel:1975}

\subsubsection{Size-dependent damping}\label{sec:kreibig}
\begin{figure}[!tb]
	\centering
	\includegraphics[width=1\columnwidth]{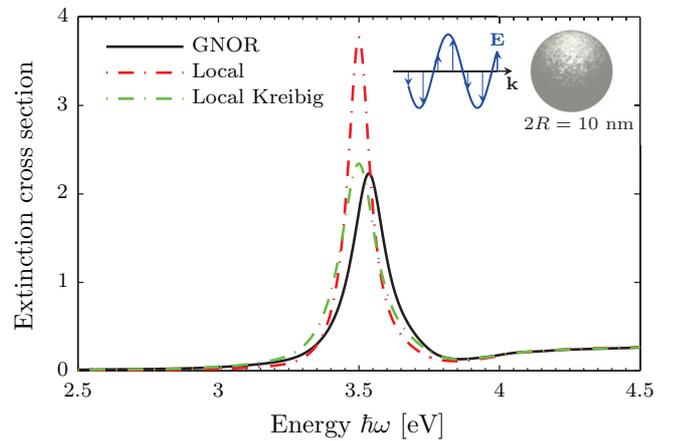}\\
	\caption{Extinction cross section (in units of the geometrical cross section $\pi R^2$ of the sphere) of a silver sphere in vacuum with diameter $2R=10$~nm as a function of energy $\hbar \omega$, calculated within the GNOR (black solid line), LRA (red dash-dotted line), and Kreibig (green dash-dotted line) approaches, see Sec.~\ref{sec:kreibig}. The value $A=0.5$ has been used for the Kreibig approach.}
	\label{fig:fig2c}
\end{figure}
The phenomenon of size-dependent damping in metal nanoparticles has been extensively observed in experiments.~\cite{Kreibig:1969,Kreibig:1985,Kreibig:1995,Gaudry:2003,Scaffardi:2006,Kolwas:2013,Genzel:1975} The theoretical approach to account for this effect in the LRA, proposed by Kreibig~\cite{Kreibig:1969} and adopted widely by researchers in the field,~\cite{Myroshnychenko:2008} has been to phenomenologically modify the Drude bulk damping parameter $\gamma$ as
\begin{equation}
	\gamma \rightarrow \gamma + A \frac{v_\textsc{f}}{R}, \label{eq:Kreibig_damping}
\end{equation}
which is only valid for spherical particles of radius $R$. Here, $A$ is a constant, which is related to the probability of the free electrons scattering off the surface of the particle. Experimental observations and advanced theoretical calculations have been compared to this approach, resulting in most cases in a value for $A$ close to unity.~\cite{Kreibig:1995,Apell:1983,Ljungbert:1983,Apell:1984} In the following, we denote the method described in Eq.~\eqref{eq:Kreibig_damping} as the Kreibig approach. In Fig.~\ref{fig:fig2b}, we compare the extinction cross section of a silver sphere with diameter $2R=10$~nm, calculated within the GNOR (black line), LRA (red line), and Kreibig (green line) approaches. As we see from Fig.~\ref{fig:fig2b} the Kreibig approach displays a size-dependent broadening as in the GNOR theory (in contrast to the LRA), but no size-dependent resonance shift (in agreement with the LRA). The SP linewidths are practically the same in the GNOR and Kreibig calculations due to the chosen value for $D$.

The Kreibig approach has been quite successful in describing the increased linewidth of SPs for different metals and geometries using a value for $A$ that is close to unity. It is therefore desirable to ensure that the value for the diffusion constant $D$ in the GNOR model gives rise to the same SP linewidth broadening as with the Kreibig approach. An estimate for $D$ can be given by comparing the size-dependent term in Eq.~\eqref{eq:omega_imag} with the size-dependent term using the Kreibig approach, i.e., $\omega_\textsc{k}'' \simeq -\gamma/2 - Av_\textsc{f}/(2R)$. Here we find the simple relation
\begin{equation}
	D=\frac{3\sqrt{10}}{5} A \frac{v_\textsc{f}^2}{\omega_\text{p}} \propto A n_0^{\frac 16}, \label{eq:D_est}
\end{equation}
which shows a linear scaling with the Kreibig parameter $A$ and a weak scaling with the equilibrium electron density $n_0$. Using appropriate material values, Eq.~\eqref{eq:D_est} provides a quite accurate estimate for $D$ (to within a factor of 2 for simple metals) compared to more thorough numerical investigations~\cite{Raza:2014b} for the plasmonic metals listed in Table~\ref{tab:tab1}. However, due to the lack of inclusion of the bound electrons (i.e., $\varepsilon_\infty \neq 1$) in Eq.~\eqref{eq:D_est}, metals with similar plasma frequency and Fermi velocity, such as gold and silver, give rise to the same value for $D$, which is only approximately correct. We have therefore used a numerical routine described by Raza~\cite{Raza:2014b} to ensure that the value for $D$ in the GNOR model gives rise to identical SP linewidth broadening for a spherical particle as the Kreibig approach. The values for $D$ (appropriate for $A=0.5$ and $A=1$) using the numerical routine are summarized in Table~\ref{tab:tab1} for several plasmonic metals.
\begin{table}[!b]
	\centering
	\caption{Plasma frequencies $\omega_\textsc{p}$, Drude damping rates $\gamma$, Fermi velocities $v_\textsc{f}$ and diffusion constants $D$ for the metals Na, Ag, Au and Al. The method used for determining $D$ is described by Raza.~\cite{Raza:2014b}}
	\label{tab:tab1}
	\begin{tabular}{lccccc}
		\hline
		& $\hbar \omega_\textsc{p}$ [eV] & $\hbar \gamma$ [eV] & $v_\textsc{f}$ [$10^6$~ms$^{-1}$] & \multicolumn{2}{c}{$D$ [$10^{-4}$~m$^2$s$^{-1}$]} \\
		\cline{5-6}
		& & & & $A=0.5$ & $A=1$ \\
		\cline{1-6}
		Na & 6.04 & 0.16 & $1.07$ & $1.08$ & $2.67$ \\
		Ag & 8.99 & 0.025 & $1.39$ & $3.61$ & $9.62$ \\
		Au & 9.02 & 0.071 & $1.39$ & $1.90$ & $8.62$ \\
		Al & 15.8 & 0.6 & $2.03$ & $1.86$ & $4.59$ \\
		\hline
	\end{tabular}
\end{table}

From Eq.~\eqref{eq:Kreibig_damping}, it is clear that the Kreibig approach is only a correction to first order in $1/R$. In contrast, the GNOR model contains corrections of order $1/R$, $1/R^2$, and so on. Thus, a signature of the GNOR model could be to find the linewidth broadening of the SP resonance to exceed the $1/R$-dependency given by the Kreibig approach. Such measurements could also be used to determine the appropriate value for $D$. Finally, we stress that size-dependent damping in non-spherical particles is also described by the GNOR model.

\subsubsection{Retardation effects}\label{sec:retardation}
In the previous sections, we have discussed the nonlocal optical response of a metal spherical particle in the nonretarded approximation, thus neglecting retardation effects. Here, we discuss the validity of the nonretarded approximation and the importance of retardation effects for different metals. One of the advantages of describing nonlocal response in the hydrodynamic approach is the ability to take into account retardation effects (in contrast to ab initio calculations such as DFT).~\cite{Eguiluz:1976,Boardman:1982a} As an example, the retarded multipolar response of a sphere including nonlocal response has been determined by Ruppin~\cite{Ruppin:1973,Ruppin:1975} by extending Mie theory to take into account longitudinal waves. In the retarded framework, the extinction cross section of a sphere in a homogeneous background can be determined using the relation~\cite{Bohren:1983,Ruppin:1973,Ruppin:1975}
\begin{equation}
	C_\text{ext} = -\frac{2\pi}{k_\textsc{b}^2} \sum_{l=1}^{\infty} \left(2l+1\right) \text{Re} \left(t_l^\textsc{te} + t_l^\textsc{tm} \right), \label{eq:extinction_ret}
\end{equation}
where $l$ denotes the angular momentum. Here, the nonlocal Mie scattering coefficients are given as~\cite{Ruppin:1973,Ruppin:1975,David:2011,Christensen:2014}
\begin{subequations} \label{eq:Mie_coefs}
	\begin{align}
	t_l^{\textsc{te}} &= \frac
	{-j_l(x_{\textsc{t}})[x_{\textsc{b}}j_l(x_{\textsc{b}})]'+j_l(x_{\textsc{b}})[x_{\textsc{t}}j_l(x_{\textsc{t}})]'}
	{j_l(x_{\textsc{t}})[x_{\textsc{b}}h_l^{\scriptscriptstyle (1)}(x_{\textsc{b}})]'-h_l^{\scriptscriptstyle (1)}(x_{\textsc{b}})[x_{\textsc{t}}j_l(x_{\textsc{t}})]'}, \label{eq:Miecoefs_TE}\\
	t_l^{\textsc{tm}} &= \frac
	{-\varepsilon j_l(x_{\textsc{t}})[x_{\textsc{b}}j_l(x_{\textsc{b}})]'+\varepsilon_{\textsc{b}}j_l(x_{\textsc{b}})\big\lbrace [x_{\textsc{t}} j_l(x_{\textsc{t}})]'+\Delta_l\big\rbrace}
	{\varepsilon j_l(x_{\textsc{t}})[x_{\textsc{b}}h_l^{\scriptscriptstyle (1)}(x_{\textsc{b}})]'-\varepsilon_{\textsc{b}}h_l^{\scriptscriptstyle (1)}(x_{\textsc{b}})\big\lbrace [x_{\textsc{t}} j_l(x_{\textsc{t}})]'+\Delta_l\big\rbrace },\label{eq:Miecoefs_TM}
	\end{align}
	where $x_\textsc{b}=k_\textsc{b}R$, $x_\textsc{t}=k_\textsc{t}R$, and $h_l^{\scriptscriptstyle (1)}$ denotes the spherical Hankel function of the first kind. The nonlocal correction $\Delta_l$ to the Mie coefficients is given as
	\begin{equation}
	\Delta_l = l(l+1) j_l(x_{\textsc{t}})\frac{\varepsilon-\varepsilon_{\infty}}{\varepsilon_{\infty}}\frac{j_l(x_{\textsc{l}})}{x_{\textsc{l}}j_l'(x_{\textsc{l}})}, \label{eq:CNL}
	\end{equation}
\end{subequations}
with $x_\textsc{l}=k_\textsc{l}R$. We note that for $l=1$ (dipole mode) the nonlocal correction in Eq.~\eqref{eq:CNL} has a similar form as $\delta_\textsc{nl}$ in the nonlocal Clausius--Mossotti factor, Eq.~\eqref{eq:alphaNL}.~\cite{Raza:2013c}
\begin{figure}[!tb]
	\centering
	\includegraphics[width=1\columnwidth]{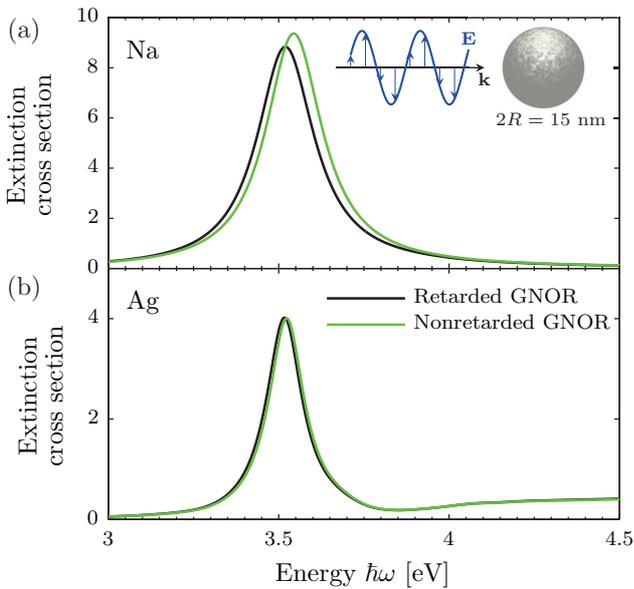}\\
	\caption{(a) Extinction cross section (in units of the geometrical cross section $\pi R^2$ of the sphere) of a sodium sphere in vacuum with diameter $2R=15$~nm as a function of energy $\hbar \omega$, calculated within the GNOR model using the fully-retarded multipolar response (black line) and the nonretarded dipolar response (green line). (b) Same as (a) but for a silver sphere.}
	\label{fig:fig2d}
\end{figure}

To examine the influence of retardation, we show in Fig.~\ref{fig:fig2d} the extinction cross section of a metal sphere in the GNOR model with a diameter $2R=15$~nm calculated with retardation (black line) and in the nonretarded limit (green line). In particular, Fig.~\ref{fig:fig2d}(a) shows the results for a sodium sphere, which is a prototypical metal studied with DFT.~\cite{Teperik:2013a,Teperik:2013b,Stella:2013,Andersen:2012,Andersen:2013} Here, we see that retardation effects already set in and give rise to a redshift of the SP resonance energy and a decrease in the extinction cross section amplitude for $R=7.5$~nm. Interestingly, this shows that retardation is important for individual nanoparticles at a size scale where nonlocal response still plays a role (cf. Fig.~\ref{fig:fig2a}).~\cite{Toscano:2014} Retardation becomes even more paramount in larger structures such as the dimer, which we discuss in Sec.~\ref{sec:dimer}. In contrast to this result on the sodium sphere, retardation shows no significant influence on a silver sphere of the same size, see Fig.~\ref{fig:fig2d}(b). For silver spheres, larger diameters must be considered ($2R>20$~nm) before retardation effects show up, which justifies our nonretarded approach in the previous sections. The validity of the nonretarded limit for the silver particle is extended to larger sizes than the sodium particle due to the permittivity of silver. This interplay between material permittivity and particle size can be clearly seen from the criterion of validity for the quasistatic approximation $\lambda \gg 2\pi R |\sqrt{\varepsilon}|$, where $\lambda$ denotes the free-space wavelength.~\cite{Bohren:1983}

\subsubsection{Multipolar response}\label{sec:multipole}
We have so far focused on nonlocal effects on the dipole mode of a small sphere and shown that this mode dominates the extinction spectrum for diameters below $20$~nm. However, other excitation sources, such as a swift electron (used in EELS measurements), produce significantly more inhomogeneous electric-field distributions than the plane-wave excitation used in determining the extinction cross section.~\cite{Christensen:2014,Abajo:2008} When in the vicinity of a metal sphere, such sources can excite higher-order multipoles, even in spheres with diameters below $20$~nm.~\cite{Christensen:2014,Abajo:1999} Thus, it is relevant to consider the nonlocal resonance condition for all multipoles given in the nonretarded limit as~\cite{Christensen:2014}
\begin{subequations}\label{eq:nl_multipole}
	\begin{equation}
	l\varepsilon + \left(l+1+\Delta_l^\text{nr}\right) \varepsilon_\textsc{b} = 0,
	\end{equation}
	where $\Delta_l^\text{nr}$ is the nonretarded limit of $\Delta_l$ of Eq.~\eqref{eq:CNL}, given as
	\begin{equation}
	\Delta_l^\text{nr} = l(l+1)\frac{\varepsilon-\varepsilon_\infty}{\varepsilon_\infty} \frac{j_l(x_\textsc{l})}{x_\textsc{l} j_l'(x_\textsc{l})}.
	\end{equation}
\end{subequations}
We see that for $l=1$, Eq.~\eqref{eq:nl_multipole} reduces to the dipole resonance condition stated in Eq.~\eqref{eq:alphaNL_pole}. As with the condition for the dipole resonance given in Eq.~\eqref{eq:alphaNL_pole}, we can determine an approximate solution to Eq.~\eqref{eq:nl_multipole} in the simple case of $\varepsilon_\textsc{b}=\varepsilon_\infty=1$. We then find the complex resonance frequencies $\omega_l=\omega_l'+i\omega_l''$ to order $1/R$ to be
\begin{subequations}\label{eq:res_multipolar}
	\begin{align}
	\omega_l' &= \frac{\omega_\text{p}}{\sqrt{1+(l+1)/l}} + \sqrt{l(l+1)} \frac{\beta}{2R}, \label{eq:res_multipolar_real} \\
	\omega_l'' &= -\frac{\gamma}{2} - \frac{l\sqrt{l+1}}{4\sqrt{2l+1}} \frac{D\omega_\text{p}}{\beta R}  \label{eq:res_multipolar_imag}
	\end{align}
\end{subequations}
which shows a clear dependence on the angular momentum $l$. Specifically, Eq.~\eqref{eq:res_multipolar} shows that the size-dependent resonance shift and linewidth broadening increases with higher order of angular momentum. This means that e.g. the quadrupolar mode of the sphere experiences a larger blueshift and linewidth broadening due to nonlocality than the dipolar mode. Higher-order modes have electric fields which are more strongly localized to the surface of the particle, and therefore are more affected by the nonlocal smearing of the induced charges. Yan~\cite{Yan:2013} provided the first theoretical observation and explanation of an $l$-dependent blueshift in the hydrodynamic model, while the $l$-dependent broadening, which is inherent for the GNOR model, is discussed here for the first time.

\begin{figure}[!tb]
	\centering
	\includegraphics[width=1\columnwidth]{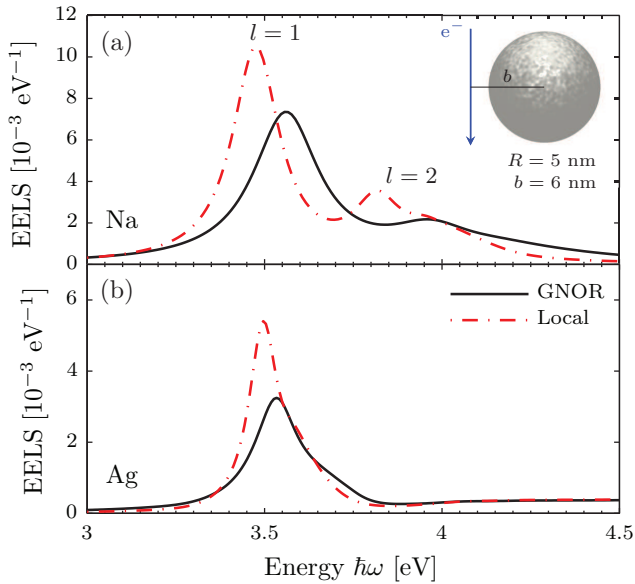}\\
	\caption{(a) Electron energy-loss probability of a sodium sphere in vacuum with diameter $2R=10$~nm as a function of energy $\hbar \omega$, calculated within the GNOR model (black line) and in the LRA (red dash-dotted line). (b) Same as (a) but for a silver sphere. The electron beam follows a straight-line trajectory with a distance of $b-R=1$~nm from the surface of the particle (see the schematic inset). The kinetic energy of the electron is $120$~keV. Calculation based on code by Christensen.~\cite{Christensen:2014}}
	\label{fig:fig2e}
\end{figure}

To exemplify the importance of higher-order modes and the $l$-dependent behavior of nonlocal response anticipated from Eq.~\eqref{eq:res_multipolar}, we consider in Fig.~\ref{fig:fig2e} the electron energy-loss (EEL) probability~\cite{Abajo:2010,Christensen:2014} of a swift electron following a straight-line trajectory near the surface of a spherical particle with diameter $2R=10$~nm. The EEL probability is directly comparable to the signal measured in EELS experiments. In Fig.~\ref{fig:fig2e}(a), we consider the EEL probability from a sodium sphere, where the low-energy peak is due to the excitation of the dipole mode ($l=1$) while the high-energy peak is due to the excitation of the quadrupole mode ($l=2$) in both the GNOR (black line) and LRA (red line) calculations. As observed in the extinction cross section, the dipole mode blueshifts and broadens in the GNOR model. Interestingly, we see that the quadrupole mode experiences a larger blueshift and linewidth broadening than the dipole mode, in accordance with our discussion from the approximate relation in Eq.~\eqref{eq:res_multipolar}. In Fig.~\ref{fig:fig2e}(b) a silver sphere is considered, where the presence of strong interband losses unfortunately dampens the higher-order modes, leaving only a resonance peak due to the excitation of the dipole mode. Other metals (in particular, Al) or settings where nonlocal response effects on higher-order plasmonic modes could be seen were discussed by Christensen.~\cite{Christensen:2014}

\subsection{Cylindrical dimer}\label{sec:dimer}
The plasmonic dimer, which consists of two metallic nanoparticles in close proximity, has attracted a lot of interest due to the plasmon hybridization~\cite{Prodan:2003,Nordlander:2004} occurring between the two closely-spaced nanoparticles, which gives rise to strongly gap-dependent resonance energies and electric-field enhancements.~\cite{Thongrattanasiri:2012} Such features have been utilized in e.g. surface-enhanced Raman spectroscopy~\cite{Kneipp:1997} and the plasmon ruler effect.~\cite{Jain:2007} The dimer has been subject to intense theoretical and experimental studies. In the simple LRA, diverging field enhancements in the gap of the dimer are encountered in the extreme case of touching dimers (i.e., no gap), which sets no limit to the number of hybridized plasmon modes,~\cite{Raza:2014b} thereby exciting an continuum of modes.~\cite{Romero:2006,Fernandez-Dominguez:2012a,Fernandez-Dominguez:2012b} These unphysical attributes of the LRA are corrected in DFT~\cite{Zuloaga:2009,Stella:2013,Teperik:2013a,Teperik:2013b,Andersen:2013} and hydrodynamic~\cite{Fernandez-Dominguez:2012a,Fernandez-Dominguez:2012b,Toscano:2012a} approaches due to the inclusion of nonlocal response and electron spill-out (only DFT). Recent measurements on dimers with subnanometer gaps using both optical techniques~\cite{Savage:2012,Cha:2014,Zhu:2014,Hajisalem:2014} and EELS~\cite{Scholl:2013,Kadkhodazadeh:2013} show lack of agreement with the LRA, and, in the touching case, also display limits on the resonance energies of the bonding plasmon modes. However, due to the indirect nature of the measurements the explanation for the discrepancy between LRA and the observed measurements is not conclusive with possible interpretations being provided from both quantum tunneling~\cite{Teperik:2013a,Teperik:2013b} and nonlocal response~\cite{Mortensen:2014} perspectives.
\begin{figure*}[!t]
	\centering
	\includegraphics[width=2\columnwidth]{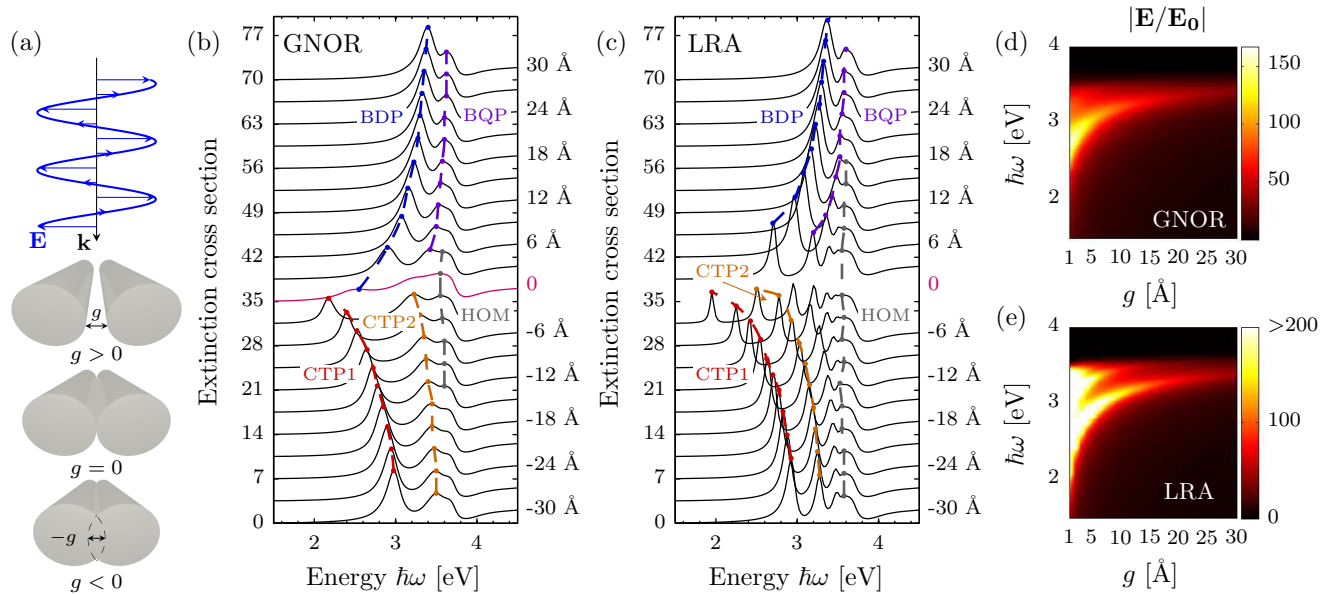}
	\caption{(a) Sketch of an incident plane wave, which is polarized along the dimer axis, impinging on a dimer consisting of two identical silver cylinders with radii $R$ and gap size $g$. (b-c) Extinction cross section (in units of the cylinder diameter $2R$) of a silver dimer in vacuum with radius $R=15$~nm for gap sizes varying from $g=30$~{\AA} (separated) to $g=-30$~{\AA} (overlapping) in steps of $3$~{\AA} calculated using the GNOR model and in the LRA, respectively. For clarity, each spectrum is displaced vertically by 3.5 normalized units. (d-e) Electric-field enhancement in the center of the dimer gap for dimers with $R=15$~nm as a function of energy $\hbar \omega$ and gap size $g$ calculated using the GNOR model and in the LRA, respectively.}
	\label{fig:fig3}
\end{figure*}

In this section, we consider a specific dimer geometry consisting of two identical silver cylinders and vary the gap from separated via touching to overlapping configurations, see Fig.~\ref{fig:fig3}(a) for a schematic illustration. The dimer consists of cylinders with radii of $R=15$~nm and is excited by a plane wave which is polarized along the dimer axis to strongly excite the hybridized modes. We numerically calculate the extinction cross section and the field enhancement in the gap of the dimer in the frameworks of the LRA and the GNOR model by using the freely available COMSOL implementation of the nonlocal equations (www.nanopl.org).~\cite{Toscano:2012a}

To clearly convey the results of the optical spectra in Figs.~\ref{fig:fig3}(b-c), we first discuss the plasmon hybridization occurring in the dimer system. When two cylinders are positioned in close proximity, their modes hybridize to form new plasmon modes, which can show up as resonances in the extinction cross section.~\cite{Nordlander:2004} For large separation distances, the first modes to hybridize are the individual dipolar modes (angular momentum $l=1$) of the cylinders to produce a lower-energy (with respect to the individual dipolar mode) bonding dipolar mode (BDP) and a higher-energy antibonding dipolar mode. Since the net dipole moment of the antibonding mode is zero, this mode will be optically dark and not show up in the extinction cross section. We therefore leave out further discussion of the antibonding modes. As the dimer separation decreases the plasmon hybridization increases and the resonance energy of the bonding dipolar mode decreases. Furthermore, with decreasing separation distance higher-order modes of the individual cylinders (i.e., $l>1$) begin to hybridize as well. As an example the quadrupole mode of the individual cylinders hybridize to form bonding and antibonding quadrupole modes. Thus, in nanometer-proximity the dimer spectra can be quite complex and show multiple resonances due to the hybridization between many modes of the individual cylinders.

With this plasmon hybridization picture in mind, we consider now in detail the extinction cross section of a silver dimer in the LRA and GNOR theory, see Figs.~\ref{fig:fig3}(b-c), respectively. We vary the dimer gap from $g=30$~{\AA} (separated dimer) via touching configuration to $g=-30$~{\AA} (overlapping dimer). We begin our discussion by considering the LRA results [Fig.~\ref{fig:fig3}(c)] for a dimer separated by a gap of $30$~{\AA}. The lowest-energy and strongest peak is due to the BDP, while the next peak is actually due to two spectrally-close modes, the bonding quadrupole mode (BQP) and a higher-order mode (HOM). Of these two modes, the BQP has the largest amplitude and lowest energy. The electric field of the HOM is concentrated at the edges of the dimer (and not in the gap, like the BDP and BQP), making it spectrally insensitive to the gap size. As the gap decreases, the BDP and BQP redshift and additional bonding-mode resonances appear due to the increased plasmon hybridization. In fact as $g\rightarrow0$ the bonding modes continue to redshift and the number of bonding modes increases without bound till a continuum of modes is found in the touching configuration $g=0$.~\cite{Fernandez-Dominguez:2012a,Fernandez-Dominguez:2012b} The extinction cross section calculation for $g=0$ does not converge in the LRA, which is why the spectrum is not present in Fig.~\ref{fig:fig3}(c). When the dimers overlap the nature of resonant modes changes and can no longer be considered as bonding modes. In particular, the induced charges pile up at the sharp edges of the junction of the overlapping dimer.~\cite{Romero:2006} The interaction between the induced charges gives rise to the several resonances seen for e.g., $g=-6$~{\AA}, which are denoted charge transfer plasmons (CTPs). As the overlap increases, the sharp edges at the junction smoothen and the interaction between the surface charges decreases, leading to a blueshift of the resonances. The spectrum of the overlapping dimer begins to increasingly resemble that of an elongated particle.

Turning our attention now to the results of the GNOR model in Fig.~\ref{fig:fig3}(b), we find that the separated dimers show less redshift with decreasing gap size compared to the results in the LRA. The size-dependent blueshift and linewidth broadening of the plasmon resonances observed in a spherical nanoparticle (see Sec.~\ref{sec:sphere}) translates into a \emph{gap}-dependent blueshift and linewidth broadening for the bonding modes of the dimer. The gap-dependent blueshift counteracts the redshift due to plasmon hybridization, leading to a finite resonance energy and finite number of bonding modes in the touching limit $g=0$ [pink curve in Fig.~\ref{fig:fig3}(b)]. Furthermore, the increase in blueshift and linewidth broadening with angular momentum observed for the spherical particle (see Sec.~\ref{sec:multipole}) manifests itself in the dimer spectra as an increased blueshift and broadening for higher-order bonding modes. Thus, the BQP experiences a stronger blueshift and broadening than the BDP, leading to a weaker plasmon hybridization for the BQP than the BDP. This is also the reason for the lower number of resonances in the GNOR spectra of close-proximity and short-overlap dimers compared to those of the LRA. For the overlapping dimers ($g<0$) the strength of the resonant modes is weaker in the GNOR calculations than in the LRA due to the nonlocal smearing of the surface charges at the geometrically sharp edges of the dimer junction.

Besides examining the extinction cross section of the dimer, we also study the electric-field enhancement present in the center of the dimer gap in Fig.~\ref{fig:fig3}(d-e). Here, we find that the GNOR model shows a significant decrease in the electric-field enhancement compared to the LRA.~\cite{Toscano:2012a} While the field enhancement in the LRA increases without bound with decreasing gap size,~\cite{Romero:2006} the GNOR model shows an amplitude- and frequency-dependence of the field enhancement in agreement with DFT simulations~\cite{Teperik:2013a,Teperik:2013b,Raza:2014b} and other models based on quantum tunneling~\cite{Esteban:2012,Haus:2014,Haus:2014b,Scalora:2014} for gaps above 5~{\AA}.

Many of the recent experimental observations on dimers in subnanometer proximity have been interpreted in a quantum tunneling framework~\cite{Savage:2012,Scholl:2013,Cha:2014} by comparing measured spectra with theoretical spectra simulated using DFT~\cite{Teperik:2013a} (or other quantum tunneling based models~\cite{Esteban:2012,Haus:2014}). But as discussed in this section, the GNOR model (which does not take into account electron spill-out) produces very similar far-field spectra (e.g., extinction cross section) as DFT-based modeling. Thus, the experimental spectra could easily also be interpreted as signatures of nonlocal response. To actually distinguish the GNOR model from quantum tunneling models (or in general theoretical models which include electron spill-out and the overlap of such), one is required to study the extreme near-field, such as the electric-field enhancement, to positively separate the effects. As far as we know, only the experimental observations of Zhu~\cite{Zhu:2014} and Hajisalem~\cite{Hajisalem:2014} have shown signs of a decrease in the electric-field enhancement when the dimer separation is smaller than 5~\AA, which seems to be in agreement with the onset of the overlap of electron spill-out.~\cite{Teperik:2013b} At these narrow gap sizes, where the overlap of electron spill-out is significant, the approximation of a hard-wall boundary condition in the GNOR model is no longer accurate, thereby setting a limit for the applicability of the model (approximately 5~{\AA} for a vacuum gap).

\subsection{Core-shell nanowire}\label{sec:coreshell}
\begin{figure}[!tb]
	\centering
	\includegraphics[width=1\columnwidth]{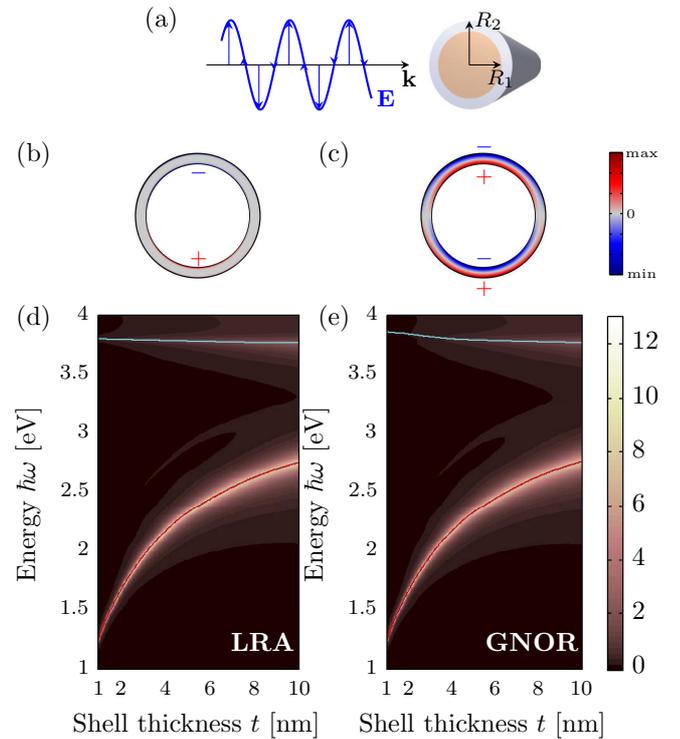}
	\caption{(a) Sketch of a plane wave impinging on a core-shell cylinder of inner radius $R_1$ and outer radius $R_2$. The permittivity of the core is $\varepsilon_\text{core}=1.5^2$, applicable for silica, and the metal shell is silver. (b-c) GNOR result of the imaginary part of the induced charge density of a silica-silver cylinder with $R_1=15$~nm and $R_2=18$~nm at the resonance energies $\hbar \omega = 1.96$~eV and $\hbar \omega=3.81$~eV corresponding to the bonding and antibonding dipole modes, respectively. (d-e) Color plots of the extinction cross section (in units of outer cylinder diameter $2 R_2$) of a silica-silver cylinder in vacuum as a function of energy $\hbar \omega$ and shell thickness $t=R_2-R_1$ calculated using the LRA and GNOR models, respectively. The inner radius is $R_1=15$~nm.}
	\label{fig:fig4}
\end{figure}

By modifying the structure of the metal nanoparticle to have a dielectric core with a metal shell, an increased tunability of the localized SP resonances (LSPRs) is achieved due to the plasmon hybridization of the inner and outer surfaces of the metal. Especially the spherical core-shell structure has received a considerable amount of attention~\cite{Brongersma:2003,Raschke:2004,Nehl:2004,Tam:2004} due to its excellent and tunable sensing properties,~\cite{Kabashin:2009} which have been utilized in biological studies such as cancer therapy.~\cite{Bardhan:2011} The plasmon hybridization allows one to position the LSPR of the nanoshell as desired by simply varying the core size $R_1$ and/or outer radius $R_2$ appropriately.~\cite{Prodan:2004}

The hybridization of the inner and outer surface plasmons increases when the metal shell becomes thinner,~\cite{Prodan:2004} which gives rise to significantly altered LSPRs compared to usual homogeneous metal nanoparticles, such as the sphere in Sec.~\ref{sec:sphere}. In Sec.~\ref{sec:dimer} we showed that the effect of nonlocal response increases with decreasing gap size (i.e., increasing hybridization). We would therefore anticipate a strong signature of nonlocal response in the core-shell particle, since it features an ultra-thin metallic shell with resulting strong plasmon hybridization.

To study the impact of nonlocal effects in the core-shell geometry, we consider an infinite cylindrical nanowire with a dielectric core and a thin metal shell excited by a plane wave, see Fig.~\ref{fig:fig4}(a) for an illustration. By extending the nonlocal Mie theory for cylinders~\cite{Ruppin:2001} to core-shell structures, we can analytically determine the extinction cross section taking into account nonlocal response in the thin metal shell and retardation effects.~\cite{Raza:2013d}

We focus on a particular design, consisting of an insulating core with permittivity $\varepsilon_\text{core}=1.5^2$ corresponding to silica, inner radius of $R_1=15$~nm, and a thin silver shell. We are interested in studying the optical response of the core-shell nanowire when varying the shell thickness $t=R_2-R_1$ by changing the outer radius $R_2$. Figure~\ref{fig:fig4}(d-e) shows the extinction cross section as a function of energy and shell thickness in the LRA and GNOR model, respectively. Considering first the LRA result, we see that the extinction cross section is dominated by two resonances: a strong low-energy resonance due to the bonding dipole mode (red line) and a weaker high-energy resonance due to the antibonding dipole mode (blue line).~\cite{Prodan:2004} The bonding mode redshifts strongly for decreasing shell thickness, while the antibonding mode blueshifts only slightly. Both shifts are due to the increase in plasmon hybridization with decreasing shell thickness. We note that a strong tunability of the bonding mode with shell thickness is present, allowing for tailoring of the optical response. When we consider the result from the GNOR theory [Fig.~\ref{fig:fig4}(e)], we find quite surprisingly the same optical response as in the LRA for the bonding mode (in stark contrast to the dimer geometry in Sec.~\ref{sec:dimer}). In particular, we see no significant size-dependent resonance shift or linewidth broadening for the bonding mode as encountered for the sphere and dimer geometries, even in the extreme case of a 1~nm thin shell. However, the antibonding mode shows an increased blueshift and a size-dependent broadening effect due to nonlocal response.

To find an answer to this surprising lack of presence of nonlocal response in the bonding mode, we consider in more detail the plasmon hybridization occurring in the core-shell geometry. In particular, we are interested in how the induced charges of the bonding and antibonding dipole modes in Fig.~\ref{fig:fig4}(d-e) are distributed.
Fig.~\ref{fig:fig4}(b) displays the induced surface charge distribution of the bonding dipole mode of the core-shell cylinder, which shows that the negative and positive induced charges are isolated to each side of the cylinder, thus separated by a distance of approximately the inner cylinder diameter, i.e., $2R_1 = 30$~nm for the design considered in Fig.~\ref{fig:fig4}. The smearing of induced charges over {\AA}ngstrom length scales due to nonlocal response will therefore not play a significant role on the optical response of the bonding plasmon mode, since the positive and negative induced charges are separated by much greater distances than the smearing length scale. In contrast, the strong effect of nonlocal response in spheres and dimers occur due to the induced positive and negative charges coming in close proximity when the particle diameter and dimer gap, respectively, are decreased. However, the antibonding mode of the core-shell geometry has induced positive and negative charges on each side of the thin metal shell [Fig.~\ref{fig:fig4}(c)], which explains why nonlocal effects play a prominent role for this mode when the metal shell is sufficiently thin.

\section{Outlook}
\subsection{Electron spill-out effect}
The nonlocal hydrodynamic and the GNOR models are the natural immediate extension to the usual Drude model for the theoretical modeling of metals. We emphasize that the difference between the nonlocal response models (that is, with hard-wall ABC and homogeneous equilibrium electron density $n_0$) and the LRA is how we model the induced charges, i.e., the charges occurring due to an exciting electric field. In the LRA the induced charges reside only on the geometric surface of the metal structure, while the inclusion of nonlocal response serves to smear out the induced surface charges on the \AA ngstrom length scale. The main shortcoming of the nonlocal models is the inaccurate treatment of the free electrons at the metal surface in the absence of an exciting electric field, i.e., the ground-state equilibrium electron density. As discussed, the free electron density is modeled as being constant inside the metal and then abruptly dropping to zero outside the metal (i.e., step profile). From the pioneering work on density-functional theory by Lang and Kohn,~\cite{Lang:1970} we know that the equilibrium electron density should be smoothly varying at the metal-vacuum interface, with Friedel oscillations inside the metal and electron spill-out just outside the metal.~\cite{Bochterle:2012} The strength of current DFT treatments of metals is the inclusion of a self-consistent treatment of the equilibrium electron density. Studying the effects of a spatially-varying electron density in the LRA has been performed,~\cite{Ozturk:2011a} however with the neglect of nonlocal gradient effects.~\cite{Ichikawa:2011} Recently, David~\cite{David:2014} and Toscano~\cite{Toscano:2014} have shown that it is also possible to properly take into account a smoothly-varying equilibrium electron density in the hydrodynamic model, thus overcoming the discussed limitations of the hard-wall nonlocal models. Expanding this approach to include diffusion should not pose significant complications, but has yet to be done.

\subsection{Metamaterials}
Metamaterials are man-made engineered materials that can manipulate and mold electromagnetic waves in ways that are not attainable with naturally available materials.~\cite{Smith:2004} A few examples of the functionalities that have been attained with metamaterials include the cloaking of macroscopic objects and negative refraction.~\cite{Shelby:2001,Ergin:2010,Chen:2011,Zhang:2011} Such properties have been designed using the theory of homogenization. Effects due to nonlocal response could come into play for optical metamaterials on two levels: firstly, in the response of the individual elements (metaatoms) as their overall sizes have to be considerably smaller than the wavelength of light, and secondly, in the arrangement of these elements at deeply subwavelength distances which enhances their nonlocal electromagnetic interaction.

While the homogenization procedure is quite complicated even in the LRA,~\cite{Pors:2011} some metamaterials have shown to be able to exhibit effects due to spatial dispersion stemming from the homogenization procedure (and not from the individual metaatoms),~\cite{Belov:2003,Silveirinha:2006,Elser:2007,Menzel:2010,Wells:2014} yet the connection to the description of the intrinsic spatial dispersion found in metals has not been made. It would be beneficial to be able to use the real-space constitutive relations derived in Secs.~\ref{sec:hydro} and \ref{sec:gnor} to also describe metamaterials with artificial spatial dispersion. This may allow a pathway for engineering spatial dispersion using metamaterials on a length scale significantly larger than the nonlocal length scale found in naturally occurring metals.

\subsection{Doped semiconductors and 2D materials}
Other free-carrier systems than metals, such as doped semiconductors~\cite{ZhangH:2014} or the 2D material graphene,~\cite{Koppens:2011,Christensen:2012} provide an alternative powerful platform to study nonlocal effects of plasmons as these materials typically allow for easier tuning of the free-carrier density $n_0$. We note that the hydrodynamic nonlocal length scale scales as (for 3D materials) $\xi \propto v_\textsc{f}/\omega_\text{p} \propto n_0^{-1/6}$, which signifies a strong dependence of nonlocal effects with the free-carrier density. While nonlocal effects in excitons in semiconductors is a well-known topic,~\cite{Pekar:1958,Hopfield:1958,Hopfield:1963,Ruppin:1981} the search for nonlocal effects in plasmons of doped semiconductors has just been initiated.~\cite{Schimpf:2014} Likewise, graphene has a $\mathbf k$-dependent dielectric function~\cite{Hwang:2007,Jablan:2009} and nonlocal effects are now being explored.~\cite{Wang:2013,Fallahi:2014,Christensen:2014b}

\subsection{Experimental ideas}
Although there have been quite some studies of surface plasmon resonances in individual nanoparticles, showing e.g. a blueshift and broadening of the LSPR with decreasing particle size, the focus has mainly been on noble metals on substrates or in inert gases. We suggest to expand this study to include nanoparticles of different metals (e.g. Al) embedded in insulating materials. For studying the size-dependent resonance behavior, EELS would be a suitable technique due to the excellent spatial and spectral resolution.

Another interesting system to see features of nonlocal response is the metallic dimer. In this system, care must be taken since \AA ngstrom-sized gaps (more precise, below 5~\AA) show signs of decreasing field enhancement in the gap of the dimer,~\cite{Zhu:2014,Hajisalem:2014} which is in contrast to expectations from the GNOR theory (cf. Sec.~\ref{sec:dimer}). However, strong nonlocal effects are still present for gap sizes above 5~\AA, where tunneling is not expected to play a role, so this makes the dimer indeed a good candidate. Additionally, the possible use of EELS to study dimers gives access to the dark modes of the system, which are also strongly gap-dependent and influenced by nonlocal response. In any case, it would also be interesting to reinterpret the existing experimental studies on dimers~\cite{Savage:2012,Scholl:2013,Cha:2014,Zhu:2014,Hajisalem:2014} using numerical calculations from the GNOR model to verify the applicability and accuracy of the nonlocal model.

The study of nonlocal effects in propagating SP modes is also of interest. In particular, continuous and homogeneous metal films support long-range SP modes~\cite{Burke:1986,Berini:2009a} which can be affected by nonlocal response for very small thicknesses.~\cite{Raza:2013b,Moreau:2013} Ultra-thin metal films supporting SP modes with long propagation lengths constitute an interesting system for measuring effects due to nonlocal response, since additional effects due to quantum tunneling, as in the dimer geometry, can be immediately ruled out.

\section{Conclusions}
We have provided a comprehensive overview of the current status of both experimental and theoretical studies on nonlocal response in plasmonic nanostructures. The real-space constitutive relations connecting the current density in metals to the electric field have been derived and discussed for the hydrodynamic and GNOR models. The GNOR model has been shown to be an extension of the hydrodynamic model by including currents due to electron diffusion. The main feature of nonlocal response is the inclusion of longitudinal waves in the metal, which gives rise to smearing of the induced charge density on the {\AA}ngstrom length scale. We have shown that, regardless of the geometry, only plasmonic modes which have negative and positive induced charges separated by a distance comparable to the smearing length scale, are affected by nonlocal response. In particular, the resonant plasmonic Mie modes supported by a spherical particle experience a blueshift and linewidth broadening as the particle diameter decreases. The same mechanism gives rise to a gap-dependent blueshift and linewidth broadening of the bonding modes of the dimer. Additionally, only the antibonding (and not the bonding) dipole mode of the core-shell geometry is affected by nonlocal response. Finally, we have presented several theoretical approaches and experimental setups to unveil and measure further effects due to nonlocal response.

\begin{acknowledgments}
We thank Thomas Christensen and Giuseppe Toscano for sharing their numerical implementations of the real-space nonlocal equations.

The Center for Nanostructured Graphene (CNG) is funded by the Danish National Research Foundation, Project DNRF58. N.~A.~M. and M.~W. acknowledge financial support by Danish Council for Independent Research--Natural Sciences, Project 1323-00087. S.~I.~B. acknowledges financial support by European Research Council, Grant 341054 (PLAQNAP).
\end{acknowledgments}


\bibliography{Raza_bib}

\end{document}